\PassOptionsToPackage{colorlinks=true,allcolors=white}{hyperref}
\documentclass[5p,times]{elsarticle}

\usepackage{amsmath} % For mathematical expressions
\usepackage{amssymb} % For mathematical symbols
\usepackage{algorithm} % For pseudocode
\usepackage{algpseudocode} % For pseudocode
\usepackage{booktabs}
\usepackage{orcidlink}
\usepackage{amsmath}
\usepackage{amssymb}
\usepackage{amsfonts}
\usepackage{multirow}
\usepackage{longtable}
\usepackage{caption}
\usepackage{subcaption}
\usepackage{pifont}
\usepackage{array}
\usepackage{url}
\usepackage{xcolor}
\usepackage{adjustbox}

\usepackage[colorlinks=true,allcolors=white]{hyperref}
\usepackage{makecell}
\usepackage{listings}
\newcolumntype{P}[1]{>{\centering\arraybackslash}p{#1}}

\begin{document}

\let\WriteBookmarks\relax
\def\floatpagepagefraction{1}
\def\textpagefraction{.001}

%\shorttitle{Raijū: Reinforcement Learning-Guided Post-Exploitation for Automating Security Assessment of Network Systems}

%\shortauthors{Van-Hau Pham et~al.}

% \title{xOffense: An AI-driven autonomous penetration testing framework with offensive knowledge-enhanced LLMs and multi agent systems}   
\title{xOffense: An Autonomous Multi-Agent Framework for Penetration Testing with Domain-Adapted Large Language Models}

%\tnotemark[1,2]

%\tnotetext[1]{This document is the results of the research
%   project funded by the National Science Foundation.}

%\tnotetext[2]{The second title footnote which is a longer text matter
%   to fill through the whole text width and overflow into
%   another line in the footnotes area of the first page.}

% tạp chí Engineering Applications of Artificial Intelligence yêu cầu nộp without author details
\affiliation[1]{organization={Information Security Lab, University of Information Technology},
    city={Ho Chi Minh City},
    country={Vietnam}}
\affiliation[2]{organization={Vietnam National University Ho Chi Minh City},
    city={Ho Chi Minh City},
    country={Vietnam}}

\author[1,2]{Phung Duc Luong \orcidlink{0009-0004-6057-5313}}   %\fnref{fn6}}
\ead{21522312@gm.uit.edu.vn}

\author[1,2]{Le Tran Gia Bao  \orcidlink{0009-0000-8911-5741}}%\fnref{fn6}}
\ead{22520105@gm.uit.edu.vn}

\author[1,2]{Nguyen Vu Khai Tam 
\orcidlink{0009-0008-1715-4213}}%\fnref{fn5}}
\ead{22521293@gm.uit.edu.vn}

\author[1,2]{Dong Huu Nguyen Khoa  \orcidlink{0009-0005-9526-140X}}%\fnref{fn5}}
\ead{23520734@gm.uit.edu.vn}

\author[1,2]{Nguyen Huu Quyen \orcidlink{0000-0002-0065-9919}}
\ead{quyennh@uit.edu.vn}

\author[1,2]{Van-Hau Pham \orcidlink{0000-0003-3147-3356}}
\ead{haupv@uit.edu.vn}

\author[1,2]{Phan The Duy \orcidlink{0000-0002-5945-3712}\corref{cor1}}%\fnref{fn2}}
\ead{duypt@uit.edu.vn}
\cortext[cor1]{Corresponding author}

\begin{abstract}

Penetration testing plays a critical role in assessing the security of modern information systems; however, existing automated approaches based on machine learning, deep learning, or reinforcement learning remain constrained by simplified action spaces, high computational overhead, and limited reasoning across multi-stage workflows such as reconnaissance, vulnerability analysis, and exploitation. Recent large language model (LLM)-based systems have shown promise in addressing these challenges, yet they often rely on large-scale or proprietary models, resulting in high cost, limited scalability, and suboptimal adaptability in complex environments.
This paper introduces xOffense, an AI-driven multi-agent framework for autonomous penetration testing that transforms traditional expert-driven processes into fully automated and scalable workflows. The proposed system leverages a fine-tuned mid-scale open-source LLM to perform structured reasoning and decision-making, while decomposing the pentesting pipeline into specialized agents responsible for reconnaissance, vulnerability scanning, and exploitation. An orchestration mechanism coordinates inter-agent collaboration to ensure coherent multi-phase execution. In addition, domain-specific fine-tuning with chain-of-thought penetration testing data enables accurate command generation and consistent multi-step reasoning across tasks. The effectiveness of xOffense is validated on two representative benchmarks, AutoPenBench and AI-Pentest-Benchmark. Experimental results demonstrate that the proposed framework consistently outperforms existing LLM-based approaches, achieving a sub-task completion rate of 79.17\% and surpassing state-of-the-art systems in both effectiveness and reliability. These results highlight that integrating domain-adapted mid-scale LLMs within a structured multi-agent architecture can provide a cost-efficient, scalable, and reproducible solution for autonomous penetration testing, offering a practical direction for deploying intelligent cybersecurity systems in real-world settings.
\end{abstract}


% Keywords

% \includegraphics{figs/grabs.pdf}

\maketitle

\section{Introduction} \label{sec_introduction}
Pentest remains one of the most effective ways to assess the real-world security posture of modern information systems. Unfortunately, the prevailing approach manual testing conducted by small teams of human experts cannot keep pace with today’s rapidly expanding attack surface. In 2024 alone, the National Vulnerability Database listed more than 29,000 new CVEs, a 38\% year-over-year increase \cite{infosecurity2024} \cite{gbhackers2024}. As networks grow in scale and complexity, the gap between the appearance of new vulnerabilities and the ability of security professionals to detect and remediate them is widening. This growing imbalance underscores the urgent necessity for automated and intelligent pentest solutions.  

Early attempts to narrow this gap focused on fully \emph{deterministic} or \emph{heuristic} automation. Notable examples include DeepExploit \cite{deepexploit2018}, and Metasploit-based scripting frameworks \cite{Metasploit} that chain banner grabbing, version mapping, and exploit invocation. Despite their efficiency, these systems rely on rigid expert rules and struggle with unseen configurations or incomplete information.  

A second, more adaptive line of work leverages \emph{RL}. Systems like IAPTS \cite{samad2024advancements} and HA-DRL \cite{tran2021deep} model pentest as a sequential decision-making problem in partially observable environments, enabling agents to autonomously explore and learn effective attack strategies through interaction and reward-based learning. RL agents can, in principle, discover novel attack paths, yet in practice they face two key obstacles: (i) their action space must be heavily simplified such as ``scan port'', ``exploit CVE-xxx'', and (ii) training requires a large number of environment interactions that are expensive to obtain and seldom transfer between real networks. Consequently, even state-of-the-art RL-based pentesters achieve modest coverage and require significant engineering to integrate new tools or protocols.  

These limitations illustrate that purely DL or RL-based automation is insufficient for the inherently multi-phase and dynamic nature of pentest. To overcome this, recent works have turned toward AI agent-based paradigms, in which multiple specialized agents collaborate to emulate the workflow of human red teams. In such systems, each agent assumes a distinct role: a \emph{Reconnaissance Agent} focuses on host and service discovery, a \emph{Vulnerability Analysis Agent} correlates findings with CVE and CWE knowledge bases, and an \emph{Exploitation Agent} generates and tests candidate payloads. This agent-oriented decomposition enables modularity, context retention across phases, and the possibility of scaling to complex attack paths that traditional ML/DL/RL pipelines cannot handle.  

Recent advancements in LLMs have opened new possibilities for automating modern pentest. Leveraging their strong reasoning and code generation capabilities, LLMs have been adopted in several research prototypes such as PentestGPT \cite{Deng2024PentestGPT}, PentestAgent \cite{shen2024pentestagent}, and VulnBot \cite{Kong2025VulnBot}, where models assist or autonomously conduct reconnaissance, scanning, and exploitation. In particular, VulnBot represents a major step forward: it frames pentest as a collaborative workflow between specialized LLM agents guided by a Penetration Task Graph (PTG), enabling the simulation of expert-level pentesting with limited or no human intervention. Empirical results from benchmarks like AutoPenBench \cite{autopenbench} and AI-Pentest-Benchmark \cite{aipentestbenchmark} have validated VulnBot’s capacity to outperform other automated methods in structured testing environments.  

However, most of these systems rely heavily on extremely large or commercial LLMs, such as GPT-4o, LLaMA3-70B, or DeepSeek-V3. Despite their capabilities, these models present significant operational hurdles including high resource consumption, costly API dependencies, and limited adaptability to domain-specific fine-tuning. Additionally, their general-purpose nature often leads to hallucinations, loss of context across phases, or poor command translation in complex penetration workflows. As such, there is a pressing need to explore whether smaller, fine-tuned open-source models can serve as more efficient, specialized alternatives, which offers better controllability, lower cost, and targeted reasoning.  

This work is motivated by two core observations. First, while large LLMs have demonstrated strong potential in security domains, scale alone does not guarantee effectiveness, particularly when models are deployed in structured, multi-step tasks like pentest. Despite their size, large-scale LLMs still suffer from context loss across phases, generate incorrect tool usage, and require significant human supervision. Second, most current systems adopt LLMs as black-box assistants or instruction followers, without integrating deeper task-specific guidance or domain adaptation. To address these limitations, we explore an alternative paradigm: leveraging a mid-sized, domain-adapted LLM that is explicitly trained for pentest tasks. We present xOffense, a refined evolution of the VulnBot framework that substitutes its dependence on large, general-purpose models with a fine-tuned Qwen3-32B \cite{yang2025qwen3}, a 32-billion-parameter open-source language model. Through dedicated training on pentest workflows, including vulnerability scanning, exploit crafting, and security tool interaction, xOffense achieves sharper task alignment, enhanced operational fidelity, and greater adaptability in nuanced or low-visibility environments.  

Beyond simply replacing the core language model, xOffense also incorporates a context-aware prompting scheme we refer to as grey-box prompting. In this setup, agents are equipped with partial system insights, such as protocol hints, observed services, or prior scan summaries, enabling them to make more informed decisions without relying on full system disclosure. This strategy preserves the operational constraints of black-box testing while offering minimal structured guidance, striking a balance between realism and agent effectiveness. By preserving VulnBot’s three-phase pipeline reconnaissance, scanning, and exploitation xOffense ensures compatibility with existing workflows, facilitates direct benchmarking, and provides a robust foundation for comparative evaluation.  

In this paper, we present the design, implementation, and evaluation of xOffense, a lightweight, domain-adaptive, and highly effective autonomous pentest system. Our key contributions are as follows:

\begin{itemize}
   \item \textbf{An AI-driven multi-agent pentest system.}  
   We propose a novel agent-based framework in which specialized agents collaborate to cover all critical phases of pentest reconnaissance, vulnerability analysis, and exploitation. This design emulates the workflow of human red teams, ensures modularity across tasks, and enables coherent orchestration of complex attack paths in an autonomous manner.  

   \item \textbf{A domain-adapted mid-scale LLM.}  
   At the core of our system lies Qwen3-32B, a 32B-parameter open-source model fine-tuned with Chain-of-Thought (CoT) pentest data. This adaptation empowers the model with precise multi-phase reasoning, accurate tool command generation, and strong adaptability in complex exploitation workflows.  

   \item \textbf{Grey-box phase prompting.}  
   We introduce a context-aware prompting mechanism that selectively integrates environmental cues such as observed protocols, discovered services, and prior scan outputs into the agent reasoning process. This strategy strikes a balance between black-box and white-box testing, reducing context loss and improving continuity across phases.  

   \item \textbf{Extensive empirical validation.}  
   We conduct rigorous evaluations of xOffense on AutoPenBench and AI-Pentest-Benchmark, demonstrating state-of-the-art performance in both synthetic and real-world pentest scenarios. The system achieves superior task and sub-task completion rates compared to prior methods, confirming the effectiveness of multi-agent orchestration and domain-adapted LLMs.  
\end{itemize}

The remainder of this paper is organized as follows. 
Section~\ref{sect_relatedworks} reviews prior studies on pentest and automation approaches, while Section~\ref{sec_background} introduces the fundamental concepts that underpin automated systems. 
Section~\ref{sec_methodology} describes the architecture of the proposed xOffense framework, including its fine-tuned Qwen3-32B model, grey-box prompting strategy, and multi-agent orchestration. We describe the experimental settings and benchmark datasets, evaluation metrics in Section~\ref{sec_experiment}. In Section~\ref{sect_evaluation_n_results} empirical results on two benchmarks and real-world exploitation scenarios are reported and analyzed. 
Section~\ref{sec_threats} discusses potential threats to validity and their implications for generalizability. In Section~\ref{sec_ethicalconsideration}, we discuss ethical considerations and responsible use of the proposed framework, including potential misuse risks, deployment constraints, and implications for defensive cybersecurity research.
Finally, Section~\ref{sec_conclusion} concludes the paper and outlines directions for future research.

\section{Related work}
\label{sect_relatedworks}

% \paragraph{Pre-LLM automation and RL}
\subsection{Pre-LLM automation and RL}
% Deterministic orchestrators (e.g., DeepExploit) integrate scanners/exploit frameworks but depend on rigid rules and shallow evidence fusion \cite{deepexploit2018,Metasploit,Nmap,Nikto,WPScan}. RL-based agents formulate pentesting as POMDP control with reward shaping \cite{tran2021deep,samad2024advancements}; while principled, deployments face two hurdles: (i) extensive state/action engineering and (ii) limited cross-target transfer without costly retraining.

Deterministic orchestrators, such as DeepExploit, integrate scanners and exploit frameworks, including Metasploit, Nmap, Nikto, and WPScan, but rely on rigid rules and shallow evidence fusion, limiting adaptability to dynamic attack scenarios \cite{deepexploit2018,Metasploit,Nmap,Nikto,WPScan}. RL-based agents formulate pentesting as a Partially Observable Markov Decision Process (POMDP) with reward shaping, providing a principled approach to automating attack strategies \cite{samad2024advancements, tran2021deep, maeda2021automating}. Notably, the Raiju framework has made significant strides in automating post-exploitation tasks by leveraging RL algorithms, specifically Advantage Actor-Critic (A2C) and Proximal Policy Optimization (PPO) \cite{pham2024raiju}. Integrated with Metasploit, Raiju trains specialized agents to perform tasks such as privilege escalation, hashdump gathering, and lateral movement in real-world environments, achieving a success rate exceeding 84\% across diverse attack types in four tested environments. However, RL-based approaches, including Raiju, face two primary challenges: (i) the need for extensive state and action space engineering, and (ii) limited cross-target transferability without costly retraining. These limitations highlight the need for more adaptive methods, such as LLMs, to enhance efficiency and generalization in pentest.

\subsection{LLM single-few-agent pipelines}
PentestGPT demonstrates a modular, self-interacting scaffold where an LLM plans, parses tool outputs, and synthesizes commands; this closes the perception$\leftrightarrow$action loop while mitigating context loss via summarization \cite{Deng2024PentestGPT}. AutoAttacker focuses on \emph{post-breach} realism with shell/Metasploit control across Windows/Linux, executing multi-step attacks \cite{Xu2024AutoAttacker}. Both highlight that language-grounded synthesis and disciplined tool use can automate substantial portions of a Cyber Kill Chain, yet often rely on large backbones and ad-hoc grounding.

\subsection{LLM multi-agent orchestration}
PentestAgent adopts RAG-grounded, role-based collaboration (reconnaissance, triage, exploitation) to reduce hallucinations and improve next step selection \cite{shen2024pentestagent}. Additionally, VulnBot structures collaboration via a PTG to preserve phase order (recon $\rightarrow$ scanning $\rightarrow$ exploitation) and constrain branching; reported results include 30.3\% overall and 69.05\% sub-task completion on AutoPenBench and strong performance on AI-Pentest-Benchmark \cite{Kong2025VulnBot,autopenbench,aipentestbenchmark}. RefPentester \cite{Dai2025RefPentester} introduces knowledge-informed self-reflection tied to stage recognition, improving recovery from failed operations on Hack The Box targets. RapidPen targets the \emph{initial foothold} (IP-to-shell) with a ReAct-style loop and retrieval of exploit knowledge, demonstrating fully autonomous compromises on HTB within minutes at modest cost \cite{Nakatani2025RapidPen}. The work of Weber et al. \cite{Weber2025Perses} presents Perses, a notable effort to enable small language models (SLMs) to perform automated privilege escalation through an extensible, role-specialized multi-LLM architecture. Perses shows that heterogeneity, which assigns lightweight models to Planner, Commander, Summariser and domain-specific Overseers, can substantially improve exploitation of simple misconfigurations. Importantly, the evaluation in Perses is narrowly scoped: experiments are conducted primarily on FreeBSD targets, employ a limited and largely handcrafted set of privilege escalation vulnerabilities, and use a threat model tailored to configuration errors rather than broad end-to-end attacks. As a result, Perses demonstrates the viability of SLM heterogeneity in constrained environments but leaves open questions about transferability to full penetration pipelines (reconnaissance, scanning, multi-stage exploitation), complex real-world services, and heterogeneous network topologies.

% The work of Weber et al. \cite{Weber2025Perses} introduces Perses, an extensible multi-LLM framework that leverages heterogeneity in model selection and task decomposition to automate privilege escalation with small backbones, aligning with cost and on-premises constraints. However, the evaluation of Perses is limited to privilege escalation, and further assessment across a broader range of pentest actions, such as lateral movement or data exfiltration, is needed to validate its generalizability.

\subsection{Focused exploit studies (one-day, zero-day) and CTF-style agents.}
Fang et al. \cite{Fang2024OneDay} show that, given CVE descriptions, GPT-4 can exploit 87\% of a 15 one-day vulnerability set, whereas other LLMs and scanners achieve 0\%; without the description, success drops markedly (7\%). Zhu et al. \cite{Zhu2025ZeroDayTeams} further extend this to teams of agents (\emph{HPTSA}) for \emph{zero-day} web vulnerabilities, reporting up to 42\% pass@5 and 18\% pass@1 on 14 real-world cases with GPT-4. For broader skill evaluation, HackSynth proposes a two-module agent and two CTF-based benchmarks (PicoCTF/OverTheWire; 200 tasks) \cite{HackSynth2024}, while NYU CTF Bench (NeurIPS D\&B) contributes a scalable open-source dataset and automation framework (200 CSAW CTF tasks) \cite{Shao2024NYUCTF}.

\subsection{Benchmarks and methodology}
The emergence of AI-driven pentest has been accompanied by a rapid proliferation of evaluation suites designed to measure autonomy, tool integration, and end-to-end performance under controlled conditions. While the space remains nascent, a few benchmarks have begun to dominate experimental protocols. Notably, \emph{AutoPenBench} and \emph{AI-Pentest-Benchmark} appear most frequently in recent studies, reflecting their alignment with realistic, multi-phase pentesting workflows and their ability to grade performance across autonomy levels and subtasks. Conversely, more specialized testbeds such as \emph{CVE-Bench} target specific exploitability dimensions, such as real-world CVEs in web contexts, and thus see adoption in works focusing on vulnerability exploitation rather than full-cycle orchestration. Capture The Flag resources such as \emph{NYU CTF Bench} and the datasets introduced by \emph{HackSynth} have also gained traction, particularly for skill-granular or task-decomposed evaluations, though their scenarios often differ from operational pentests in scope and realism.

Within this landscape, AutoPenBench offers open and standardized graded tasks spanning web, network, and cryptographic targets, with configurable autonomy modes to support comparisons between orchestration strategies and model backbones \cite{autopenbench}. AI-Pentest-Benchmark provides VM-based targets, enabling reproducible end-to-end penetration tests \cite{aipentestbenchmark}, thereby supporting performance attribution across discovery, exploitation, and post-exploitation phases. CVE-Bench grounds evaluation in real-world web CVEs, reporting typical success rates in the low teens even for state-of-the-art agents, highlighting the gap between research prototypes and robust autonomy \cite{zhu2025cvebench}. Methodological recommendations across these works increasingly emphasize standardizing budget constraints, clearly labeling autonomy levels, and reporting detailed error modes to prevent overestimation of capabilities \cite{Isozaki2025Towards}.

\subsection{Positioning of our work}
Relative to single-agent \emph{PentestGPT} \cite{Deng2024PentestGPT} and post-breach \emph{AutoAttacker} \cite{Xu2024AutoAttacker}, we retain a multi-agent/PTG discipline akin to \emph{PentestAgent}/\emph{VulnBot} \cite{shen2024pentestagent,Kong2025VulnBot} but differ in three ways: (i) prioritizing mid-scale, open backbones for cost-effective, on-prem deployment; (ii) employing \emph{grey-box phase prompting} to maintain phase continuity while effectively limiting drift; and (iii) aligning evaluation with open substrates (AutoPenBench, AI-Pentest-Benchmark, and where applicable CVE-Bench) under fixed budgets and sub-task breakdowns \cite{autopenbench,aipentestbenchmark,zhu2025cvebench,Isozaki2025Towards}.

% ================================
% Wider, grid-lined comparison table. Compact abbreviations + legend avoid overflow.
\begin{table*}[t!]
\centering
\caption{Comparison of representative automated pentesting systems and benchmarks (grid-lined). Criteria emphasize architecture, scope, grounding, tools, autonomy, and evaluation.}
\label{tab:rw_compare}
\small
\renewcommand{\arraystretch}{1.12}
\setlength{\tabcolsep}{3.8pt}
\begin{adjustbox}{width=1.0\linewidth}
\begin{tabular}{|l|c|c|c|c|c|}
\hline
\textbf{Work} & \makecell{\textbf{Architecture}\\\textbf{(Arch.)}} & \makecell{\textbf{Scope/Phase}} & \makecell{\textbf{Grounding/}\\\textbf{Memory}} & \makecell{\textbf{Tool Use}} & \makecell{\textbf{Evaluation} \\ \textbf{\& Highlights}} \\
\hline
PentestGPT \cite{Deng2024PentestGPT} & SA & Web+Net (multi-phase) & SUM (module summaries) & Parse scans $\to$ cmd synth. & USENIX'24 cases/bench; modular pipeline \\
\hline
AutoAttacker \cite{Xu2024AutoAttacker} & SA (post) & Post-breach, OpSec realism & CTX (in-session) & Shell, Metasploit & Simulated org (Win/Linux); multi-step attacks \\
\hline
PentestAgent \cite{shen2024pentestagent} & MA & Web focus (ext. assess.) & RAG + MEM & Scanners, PoCs & Bench + HTB; reduced hallucinations via RAG \\
\hline
VulnBot \cite{Kong2025VulnBot} & MA + PTG & Full cycle (recon$\to$exp.) & Phase SUM (PTG state) & Nmap, Nikto, Metasploit & AutoPenBench (30.3\% overall; 69.05\% sub-task), AIPB best-of-six \\
\hline
RefPentester \cite{Dai2025RefPentester} & MA + RFL & Stage-aware triage/exp. & Reflection + knowledge & Std.\ toolchain & HTB ``Sau'': +16.7\% vs GPT-4o baseline \\
\hline
Perses \cite{Weber2025Perses} & MA (multi-LLM) & Privilege escalation & \textbf{HET} (heterogeneous model/task) & Tool-grounded (details in paper) & FreeBSD systems; small-LLM focus \\
\hline
RapidPen \cite{Nakatani2025RapidPen} & SA & \textbf{IP$\to$Shell} (initial foothold) & MEM + exploit retrieval & Scan$\to$exploit loop & HTB: autonomous shells in minutes; low cost \\
\hline
One-Day Agent \cite{Fang2024OneDay} & SA & Web (one-day CVEs) & CVE-guided CTX & Browser, tools & 87\% (GPT-4, with CVE desc.); 7\% w/o desc. \\
\hline
HPTSA (Teams) \cite{Zhu2025ZeroDayTeams} & MA (hier./team) & Web (zero-day) & Planner + experts & Browser, task agents & 42\% pass@5; 18\% pass@1 on 14 real vulns \\
\hline
HackSynth \cite{HackSynth2024} & SA (2-mod.) & CTF (200 tasks) & Planner + summarizer & Sandbox tools & PicoCTF/OTW benchmarks; GPT-4o best \\
\hline
AutoPentest \cite{Henke2025AutoPentest} & MA (LangChain) & Black-box (enum$\to$exp.) & Prompting + MEM & Std.\ scans/exploits & GPT-4o-based prototype; open code \\
\hline
CVE-Bench \cite{zhu2025cvebench} & Bench & Real web CVEs & -- & -- & Up to 13\% s-rates for SOTA agents \\
\hline
AutoPenBench \cite{autopenbench} & Bench & Mixed (Web/Net/CRPT) & -- & -- & 33 tasks; autonomy and milestone scoring \\
\hline
NYU CTF Bench \cite{Shao2024NYUCTF} & Bench & CTF (CSAW; 200) & -- & Tool-integrated & NeurIPS D\&B; open dataset+automation \\
\hline
\textbf{Our work} & MA + PTG & Full cycle (Web+Net) & \textbf{GBP} + MEM & Broad scan/exploit & APB, AIPB, (opt.) CVEB; mid-scale open LLM \\
\hline
\end{tabular}
\end{adjustbox}

\vspace{4pt}
\footnotesize
\textbf{Legend:} SA=Single-agent; MA=Multi-agent; PTG=Penetration Task Graph; RFL=Reflection; SUM=Summaries; RAG=Retrieval-Augmented Generation; MEM=Explicit memory; CTX=In-session context; \textbf{HET}=Heterogeneity (multi-LLM model/task selection); GBP=Grey-box phase prompting; APB=AutoPenBench \cite{autopenbench}; AIPB=AI-Pentest-Benchmark \cite{aipentestbenchmark}; CVEB=CVE-Bench \cite{zhu2025cvebench}. ``s-rate'' denotes success rate.

\end{table*}

\subsection{Takeaways.}
Outcomes hinge on (i) \emph{grounding quality} (RAG/summaries/validators) and (ii) \emph{orchestration discipline} (roles/PTG/reflection). Single-agent pipelines set baselines; role-structured multi-agent systems consistently improve reliability; reflective and IP-to-shell variants further push autonomy at kill-chain ends; one-day/zero-day studies quantify limits of discovery vs.\ exploitation \cite{Fang2024OneDay,Zhu2025ZeroDayTeams}. Our approach emphasizes cost-effective, reproducible deployment with mid-scale open models, PTG structure, and grey-box prompts under open, reproducible protocols \cite{autopenbench,aipentestbenchmark,zhu2025cvebench,Isozaki2025Towards}.
\section{Background} \label{sec_background}

\subsection{Automated Pentest}

Pentest aims to evaluate the security of a target system by simulating adversarial behavior across phases such as reconnaissance, vulnerability enumeration, exploitation, and privilege escalation. Manual execution is effective but limited by human resources and scalability. Automated Pentest (APT) addresses these limitations by orchestrating these phases through intelligent agents and machine learning models.

Formally, let $T$ denote a target system with configuration space $\mathcal{C}$ and attack surface $\mathcal{S}$. An automated pentesting operation can be represented as a pipeline:
\[
f: T \mapsto \{R, V, E, P\} \mapsto O
\]
where $R$ are the reconnaissance results (asset discovery, service mapping), $V$ are detected vulnerabilities, $E$ denotes the exploit simulation results, $P$ represents the privilege escalation attempts and $O$ is the structured output (reports, risk scores, or attack paths). 

Such pipelines resemble MLOps workflows, where data collection, model inference, and result verification are continuously orchestrated. In this analogy, reconnaissance and vulnerability scanning serve as data ingestion, exploitation is model inference, and reporting acts as the 'evaluation' stage. By automating this operation, APT enables repeatability, scalability, and integration into CI/CD security pipelines.

\subsection{Multi-Agent AI Systems}

Multi-Agent Systems (MAS) provide a natural architecture for automated pentest by assigning each pentest phase to a specialized agent. For example:
\begin{itemize}
    \item Reconnaissance Agent: enumerates hosts, ports, and services (similar to VulnBot's 'Recon Agent' \cite{Kong2025VulnBot}).
    \item Vulnerability Analysis Agent: correlated scan data with CVE / CWE knowledge bases.
    \item Exploitation Agent: generates and tests candidate payloads 
    \item Reporting Agent: summarizes results, attack graphs, and remediation advice.
\end{itemize}

Agents communicate through a task manager or memory module, allowing modularity and fault tolerance. Frameworks like \textit{CAMEL} \cite{NEURIPS2023_a3621ee9} show that role-conditioned LLM agents can collaborate effectively on complex objectives. In our system \textit{xOffense}, MAS design ensures that each offensive task is handled by a role-specialized model while maintaining global coordination.

\subsection{LLM-based Offensive Agents}

Within MAS, LLMs provide reasoning, contextual understanding, and code synthesis capabilities that align well with pentest workflows. The central problem is, given an attack context $C$ (system description, logs, CVEs), generate actionable steps or payloads $A$ that maximize the likelihood of successful exploitation:
\[
g: C \mapsto A
\]

Although proprietary LLMs, such as GPT-4 and Claude, offer strong performance, they introduce limitations in cost, reproducibility, and security control. Therefore, we adopt open-source models such as \textbf{Qwen3}, which support local deployment, fine-tuning, and quantization (AWQ/INT4), making them suitable for offensive research environments where sensitive data cannot leave the infrastructure.

To mitigate risks such as hallucination or unsafe outputs, LLM agents are sandboxed and validated against controlled execution environments before results are accepted.

\subsection{Fine-tuning Methods for Offensive LLMs}

Adapting LLMs to offensive security tasks requires task-specific specialization beyond general pre-trained knowledge, and several parameter-efficient fine-tuning approaches have been  explored, including Prefix-Tuning, Adapter-based methods, Low-Rank Adaptation (LoRA) and its extension QLoRA.  

Prefix-Tuning or P-Tuning v2 appends task-specific continuous vectors or tokens to the input prompt. Its advantages lie in simplicity and a very low memory footprint, since only the prefix embeddings are optimized. However, this method often struggles to capture deeper structural knowledge, and its effectiveness diminishes when reasoning requires multi-step tool interaction or long-context planning, both of which are essential in pentest.  

Adapter-based methods insert small trainable modules within transformer layers, enabling modular adaptation to new tasks. They provide good task isolation and make it possible to reuse the same backbone across different domains with minimal additional parameters. Nonetheless, these methods introduce extra inference latency due to the added modules, and their limited capacity makes them less effective for embedding highly domain-specific procedural knowledge such as exploit reasoning or vulnerability chaining.  

LoRA and its extension QLoRA decompose weight updates into low-rank matrices, significantly reducing the number of trainable parameters while retaining the expressive power of the base model. LoRA achieves a balance between efficiency and performance: it requires far fewer resources than full fine-tuning, adds negligible inference overhead, and can effectively embed specialized knowledge without erasing the general reasoning ability of the original model. Conceptually, weight updates $\Delta W$ lie in a low-rank subspace:
\[
\Delta W = A B, \quad A \in \mathbb{R}^{d \times r}, \; B \in \mathbb{R}^{r \times k}, \; r \ll \min(d,k)
\]
and the effective weight during inference is
\[
W' = W + \alpha A B
\]
where $\alpha$ is a scaling factor controlling the magnitude of the adaptation.  

Considering the need to adapt a 32B-parameter model such as Qwen3 under realistic hardware constraints, this work adopts LoRA fine-tuning. This approach makes it possible to embed offensive knowledge, including vulnerability patterns, exploit reasoning, and payload generation, efficiently while preserving the base model’s general-purpose capabilities. Detailed dataset construction and the training procedure are described in Section~\ref{sec_methodology}.

% \subsection{Fine-tuning Methods for Offensive LLMs}

% Adapting LLMs to offensive security tasks requires specialization beyond pre-trained general knowledge. Several parameter-efficient fine-tuning methods exist:
% \begin{itemize}
%     \item Prefix-Tuning / P-Tuning v2: injects task-specific tokens into prompts.
%     \item Adapter-based Methods: adds small task-specific modules to transformer layers.
%     \item LoRA / QLoRA: factorizes weight updates into low-rank matrices, drastically reducing trainable parameters.
% \end{itemize}

% Among these, LoRA offers the best trade-off between efficiency and performance, allowing large models like Qwen3 to be adapted on modest hardware while preserving general knowledge. Conceptually, weight updates $\Delta W$ lie in a low-rank subspace:
% \[
% \Delta W = A B, \quad A \in \mathbb{R}^{d \times r}, \; B \in \mathbb{R}^{r \times k}, \; r \ll \min(d,k)
% \]
% During the inference, the effective weight is the following:
% \[
% W' = W + \alpha A B
% \]
% where $\alpha$ is a scaling factor.  

% For \textit{xOffense}, we employ LoRA fine-tuning to embed offensive knowledge such as vulnerability patterns, exploit reasoning, and payload generation. Detailed dataset construction and training methodology are presented in Section~\ref{sec_methodology}.
% \input{sections/2.1-problem-definition}
\section{Methodology} \label{sec_methodology}
\subsection{Overview of the proposed framework}
xOffense is an innovative, lightweight framework for autonomous pentest, engineered to replicate the collaborative dynamics of human security teams while operating within resource-constrained environments. By harnessing compact LLMs with approximately 32 billion parameters, xOffense eliminates dependency on commercial APIs, enabling deployment on standard hardware. The framework decomposes the intricate process of pentest into three meticulously designed phases reconnaissance, scanning, and exploitation coordinated through a sophisticated multi-agent architecture. Comprising five core components Task Orchestrator, Knowledge Repository, Command Synthesizer, Action Executor, and Information Aggregator. xOffense ensures seamless task progression, robust information management, and precise execution. This section elucidates the system’s architecture, role delineation, task coordination, inter-agent communication, and execution mechanisms, underscoring its efficacy in addressing cybersecurity challenges with minimal computational overhead.

\begin{figure*}[h!]
    \centering
    \includegraphics[width=\textwidth]{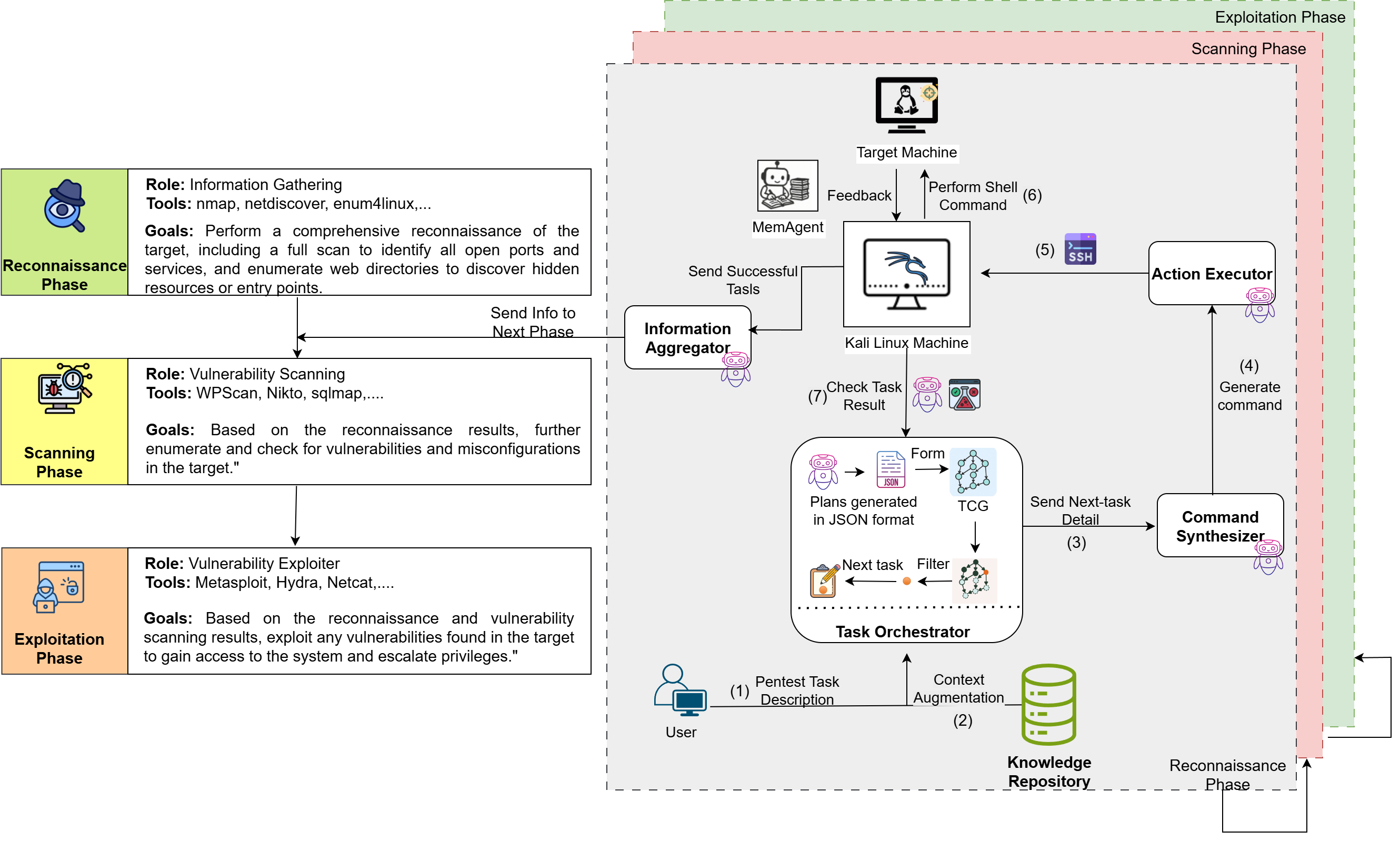}
    \caption{The Overall Architecture of the xOffense Framework.}
    \label{fig:pipeline}
\end{figure*}

The operational workflow of \textit{xOffense}, illustrated in Figure \ref{fig:pipeline}, initiates when a user submits a pentest objective, such as ``Identify vulnerabilities on IP 192.168.X.X and retrieve root-level flags.'' This task description serves as the Initial Context, which is passed to the \textit{Task Orchestrator} for comprehensive plan generation. The orchestrator constructs a Task Coordination Graph (TCG), decomposing the penetration objective into a structured sequence of tasks with clearly defined dependencies. To enhance contextual accuracy, it queries the \textit{Knowledge Repository} a vector-based database via a Retrieval-Augmented Generation (RAG) mechanism, which is taken from Langchain-Chatchat \cite{langchain_chatchat}, retrieving relevant penetration knowledge based on inputs such as the initial task description, the current task's instruction, or recent task results.

Each task within the TCG is processed through an iterative and adaptive loop involving Command Synthesis, Execution, Feedback Analysis, and Dynamic Plan Update. The \textit{Command Synthesizer}, fine-tuned using lightweight LoRA techniques, translates task directives into precise, tool-specific commands, which are executed by the \textit{Action Executor} utilizing a MemAgent-enhanced context management system to handle verbose outputs effectively. Post-execution, task outcomes are evaluated and relayed back to the orchestrator, which marks tasks as completed or triggers reflection and re-planning in case of failures. Upon completing all tasks within a phase, such as \textit{Reconnaissance}, the \textit{Information Aggregator} consolidates outputs into concise directives for the subsequent phase, such as \textit{Scanning}, ensuring contextual coherence and minimizing token overhead. This orchestration-execution-feedback loop is consistently applied across all three phases, with each phase iteratively refining the attack path based on task outcomes and environmental feedback. The overall workflow, including execution and re-planning, is later formalized in Algorithm~\ref{alg:check-reflection}. The workflow terminates upon successfully achieving the defined success criteria, such as privilege escalation or flag retrieval.

\subsection{Role Specialization}
To navigate the complexity of pentest, xOffense employs a role specialization strategy, mitigating the risk of information overload and ensuring contextual coherence across phases. By assigning agents to distinct roles, the framework optimizes resource utilization and maintains precision in task execution, addressing the challenge of dynamic reasoning across testing stages.

\begin{itemize}
    \item \textbf{Reconnaissance Phase}: Agents focus on comprehensive intelligence gathering, cataloging network configurations, open ports, and service details. Tools such as \textit{Nmap} \cite{Nmap} for network scanning, \textit{Dirb} \cite{Dirb} for directory enumeration, \textit{Gobuster} \cite{Gobuster} for brute-forcing hidden directories, and \textit{Amass} \cite{Amass} for subdomain discovery are integrated. For instance, a task might execute \texttt{nmap -sV -p- <target-ip>} to map all open ports and services, providing a robust foundation for subsequent phases.

    \item \textbf{Scanning Phase}: Building on reconnaissance insights, scanning agents identify vulnerabilities and misconfigurations using tools like \textit{Nikto} \cite{Nikto} for web server analysis, \textit{WPScan} \cite{WPScan} for WordPress vulnerabilities, \textit{sqlmap} \cite{sqlmap} for SQL injection testing for comprehensive vulnerability scanning. This phase prioritizes exploitable weaknesses to streamline progression.

    \item \textbf{Exploitation Phase}: Agents exploit identified vulnerabilities to gain unauthorized access or escalate privileges, employing tools such as \textit{Metasploit} \cite{Metasploit} for exploit development, \textit{Hydra} \cite{Hydra} for credential brute-forcing, \textit{John the Ripper} \cite{JohnTheRipper} for password cracking, and \textit{ExploitDB} \cite{ExploitDB} for sourcing exploit scripts. For example, a task might deploy a Metasploit module to exploit a known CVE, followed by privilege escalation via a custom script.
\end{itemize}

This structured delineation ensures that each phase leverages prior findings, fostering a cohesive testing process and mitigating the risk of fragmented analyses.

\subsection{Task Coordination and Reflection}

The TCG and its integrated \textit{Check and Reflection Mechanism} form the cornerstone of xOffense's penetration path planning, enabling systematic task execution and adaptive plan refinement. These components address the challenges of limited context windows and inadequate error handling, ensuring robust and dynamic testing workflows. Their operational logic is illustrated in Algorithm~\ref{alg:check-reflection}, Algorithm~\ref{alg:updateplan}, and Algorithm~\ref{alg:mergetasks}.

\subsubsection{Task Coordination Graph}

The TCG is a structured acyclic digraph, defined as $G = (V, E)$, where $V$ represents individual tasks and $E$ denotes dependencies, ensuring logical and conflict-free execution. Each task node $v \in V$ encapsulates attributes such as:

\begin{itemize}
    \item Directive: A clear instruction, such as \textit{"enumerate services on port 80 of 192.168.X.X."}
    \item Operation Type: Specifies whether the task involves automated shell commands, such as \texttt{nmap}, or manual intervention.
    \item Prerequisites: Lists tasks that must be completed prior to execution, ensuring sequential integrity.
    \item Command: The tool-specific instruction generated by the \textit{Command Synthesizer}.
    \item Outcome: The execution result, capturing tool outputs or errors.
    \item Completion Status: Indicates whether the task is completed or pending.
    \item Success Status: Records whether the task was successful.
\end{itemize}

The \textit{Task Orchestrator} generates the TCG in a JSON, which is compliant format, dynamically updating it based on execution outcomes. For instance, a task designed to perform user authentication by initiating an SSH connection attempt, targeting a specific service endpoint associated with remote shell access protocols, on port 22 of 192.168.X.X depends on the successful completion of a preceding port scanning task. Subsequent tasks, such as performing an exhaustive enumeration of writable directories for privilege escalation through misconfigured permissions or publicly writable paths (\texttt{find / -writable 2>/dev/null}) or listing running processes (\texttt{ps aux}), are contingent on this authentication.

\begin{figure*}
    \centering
    \includegraphics[width=0.85\linewidth]{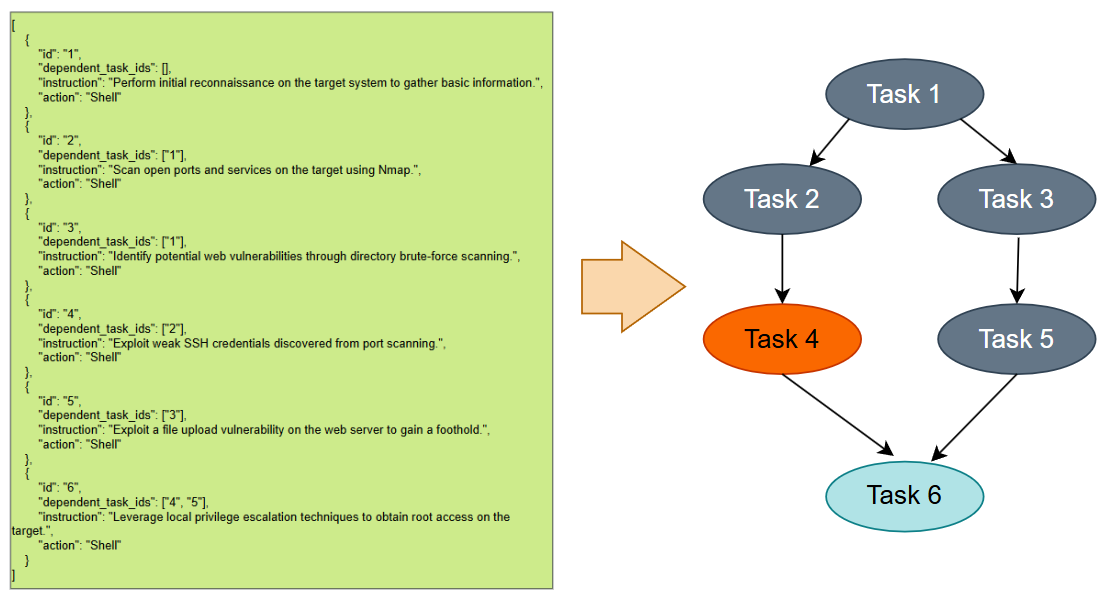}
    \caption{\textbf{TCG} illustrating task dependencies and execution status. Completed tasks are shown in dark, the current task in orange, and pending tasks in light blue.}
    \label{fig:tcg}
\end{figure*}

Figure~\ref{fig:tcg} illustrates a sample TCG, with a JSON task list on the left detailing directives, dependencies, and commands, and a dependency graph on the right showing task sequences with arrows indicating prerequisites. This formalism provides the structural foundation later used in Algorithm~\ref{alg:check-reflection} for iterative execution and feedback handling.

The TCG operates through two sessions:

\begin{itemize}
    \item \textbf{Planning Session}: The \textit{Task Orchestrator} constructs an initial action plan tailored to the target system’s characteristics and user requirements. It decomposes the plan into structured task lists, ensuring logical sequencing and dependency alignment. The plan is dynamically refined based on execution feedback, addressing the challenge of maintaining coherent context across phases.
    
    \item \textbf{Task Session}: This session generates detailed instructions for each task, which are passed to the \textit{Command Synthesizer} for command generation and to the \textit{Action Executor} for execution. It also evaluates execution outcomes, updating the TCG’s completion and success statuses, as shown in Algorithm~\ref{alg:check-reflection}.
\end{itemize}

This structured approach ensures systematic progression, mitigating the risk of out-of-sequence execution and enhancing efficiency in resource-constrained environments.

\subsubsection{Check and Reflection Mechanism}
The generation of erroneous commands and the lack of effective error-handling mechanisms pose significant challenges to LLM-based pentest. xOffense addresses these issues through a \textit{Check and Reflection Mechanism} integrated into the \textit{Task Orchestrator}, enabling continuous self-assessment and plan optimization. The full workflow is detailed in Algorithm~\ref{alg:check-reflection}.

During the \textit{Task Session}, the \textit{Action Executor} evaluates task outcomes and updates the \textit{TCG} with success or failure statuses. The \textit{Planning Session} then reflects on these outcomes, revising task directives and updating the TCG accordingly. Successful tasks are retained, while failed tasks trigger a reanalysis process, wherein the LLM regenerates commands with corrected parameters or alternative strategies. The updated plan is merged with previously completed tasks to preserve execution continuity, as demonstrated in Algorithm~\ref{alg:updateplan} and Algorithm~\ref{alg:mergetasks}.

\begin{algorithm}[H]
\caption{Check and Reflection Procedure}
\label{alg:check-reflection}
\begin{algorithmic}[1]
\Require TCG, Knowledge Repository $KR$
\While{not all tasks completed}
    \State $t \gets$ \texttt{NextTask}(TCG)
    \State $r \gets$ \texttt{Execute}(t)
    \If{\texttt{CheckSuccess}($r$)}
        \State \texttt{MarkCompleted}($t$)
        \State \texttt{StoreEmbedding}($t$, $r$, $KR$)
    \Else
        \State $K \gets$ \texttt{RetrieveSimilar}(t, $KR$)
        \State $t' \gets$ \texttt{RegenerateTask}(t, $K$)
        \State \texttt{MergeTasks}(t', TCG)
    \EndIf
    \State \texttt{UpdatePlan}(TCG)
\EndWhile
\end{algorithmic}
\end{algorithm}

\begin{algorithm}[H]
\caption{\texttt{UpdatePlan}: LLM-driven Plan Revision}
\label{alg:updateplan}
\begin{algorithmic}[1]
\Require Current plan $P$, failed task $t$, result $r$
\State $S \gets$ list of completed-success tasks from $P$
\State $F \gets$ list of failed tasks from $P$
\State $P_{\text{new}} \gets$ \textsc{LLMUpdatePlan}($t$, $r$, $S$, $F$)
\State $P^\star \gets$ \texttt{MergeTasks}($P$, $P_{\text{new}}$)
\State \Return $P^\star$
\end{algorithmic}
\end{algorithm}

\paragraph{Plan Update and Merge Algorithms.}
The \texttt{UpdatePlan} and \texttt{MergeTasks} procedures are key to preserving execution continuity. Upon task failure, the system calls the LLM to propose an updated plan, then merges it with the existing TCG such that all successfully completed tasks are retained, sequence numbers are adjusted, and only pending or failed tasks are revised. Their formal pseudocode is shown in Algorithm~\ref{alg:updateplan} and Algorithm~\ref{alg:mergetasks}.

The \textit{Knowledge Repository} supports this mechanism indirectly by assisting the \textit{Task Orchestrator} during plan updates. It stores embeddings of previously successful tasks and curated pentest knowledge, including exploitation techniques, privilege escalation methods, and tool usage tutorials from sources such as HackTricks~\cite{HackTricks} and HackingArticles~\cite{HackingArticles}. When re-planning, the \textit{Task Orchestrator} queries this repository to retrieve the top-$k$ most relevant past cases using vector similarity search. Retrieved results are re-ranked and integrated into the revised TCG, ensuring that updates benefit from prior successes. This integration enhances resilience against hallucinated commands, improves error recovery, and maintains efficiency across iterative pentest phases.

\begin{algorithm}[H]
\caption{\texttt{MergeTasks}: Success-Preserving Integration}
\label{alg:mergetasks}
\begin{algorithmic}[1]
\Require Old plan $P_{\text{old}}$, new plan $P_{\text{new}}$
\State $\mathcal{C} \gets$ completed-success tasks from $P_{\text{old}}$
\State $\mathcal{M} \gets$ empty list
\ForAll{$\tau \in \mathcal{C}$}
    \If{$\tau$ not in $P_{\text{new}}$ by instruction}
        \State append $\tau$ to $\mathcal{M}$ with reset dependencies
    \EndIf
\EndFor
\ForAll{$\hat{\tau} \in P_{\text{new}}$}
    \If{instruction matches a task in $\mathcal{C}$}
        \State reuse completed task with updated dependencies
    \Else
        \State append $\hat{\tau}$ as a new task
    \EndIf
\EndFor
\State update sequence numbers in $\mathcal{M}$
\State \Return merged plan with tasks $\mathcal{M}$
\end{algorithmic}
\end{algorithm}

\subsection{Inter-Agent Communication Mechanism}

Seamless coordination among agents is critical for maintaining contextual coherence across the reconnaissance, scanning, and exploitation phases, particularly given the limited context window of compact LLMs. xOffense employs the \textit{Information Aggregator} to facilitate efficient communication, consolidating verbose outputs into concise, actionable summaries to optimize token usage and prevent information overload. The entire communication pipeline is operationalized in Algorithm~\ref{alg:inter-agent}.

For example, reconnaissance outputs, such as open ports (such as 22, 80, and 443), service versions, and system fingerprints, are synthesized into a compact directive for the scanning phase, enabling targeted vulnerability detection with tools like \textit{Nikto} or \textit{sqlmap}. Similarly, scanning outputs, such as a SQL injection vulnerability identified by \textit{sqlmap} or a misconfiguration detected by \textit{Nuclei}, are summarized to guide the exploitation phase in prioritizing relevant exploits.

The \textit{Information Aggregator} maintains a persistent shell state log, tracking access levels, such as a low-privileged user account gained via SSH, and system context, such as operating system type. This log ensures continuity across phases, mitigating the risk of context loss and enabling dynamic integration of findings. By filtering outputs from preceding phases to focus on critical insights, the mechanism minimizes computational overhead, ensuring efficient operation on a 32B-parameter LLM. This streamlined communication fosters a cohesive testing process, addressing the challenge of synthesizing information across multiple stages.

\setlength{\textfloatsep}{8pt plus 2pt minus 2pt}
\setlength{\intextsep}{8pt plus 2pt minus 2pt}
\setlength{\floatsep}{8pt plus 2pt minus 2pt}

% ... phần prose giữ nguyên ở trên ...

\begin{algorithm}[t] % đổi [H] -> [t] để tránh “ô trắng”, cho float lên đầu trang
\caption{Inter-Agent Communication via PlannerSummary}
\label{alg:inter-agent}
\begin{small} % thu nhỏ nhẹ để khối gọn hơn, dễ “fit” một trang
\begin{algorithmic}[1]
\State \textbf{Input:} Phase sequence $P = \{p_1, p_2, \dots, p_n\}$, Shell State Log $S$
\For{$i = 1$ to $n-1$}
    \State \textbf{// Step 1: Collect and summarize results from previous phase}
    \State $\texttt{history\_ids} \gets \texttt{GetPlannerIDs}(p_i)$
    \If{$|\texttt{history\_ids}| = 0$}
        \State $context \gets ""$
    \Else
        \State $summary \gets$ \texttt{"Previous Phase:\textbackslash n"}
        \For{each $id$ in $\texttt{history\_ids}$}
            \State $plan \gets \texttt{get\_planner\_by\_id}(id)$
            \For{each $task$ in $plan.\texttt{finished\_tasks}$}
                \State $summary \gets summary$ \texttt{ \string|\string| "Instruction: "} $+\;task.\texttt{instruction}$
                \Statex \hspace{\algorithmicindent}$\;\;\;+\,$ \texttt{ \string|\string| "Code: "} $+\;task.\texttt{code}$
                \Statex \hspace{\algorithmicindent}$\;\;\;+\,$ \texttt{ \string|\string| "Result: "} $+\;task.\texttt{result}$
                \Statex \hspace{\algorithmicindent}$\;\;\;+\,$ \texttt{"\textbackslash n  \textbackslash n"}
            \EndFor
        \EndFor
        \State $context, \_ \gets \texttt{callLLM}(\texttt{query}=\texttt{write\_summary} + summary,\; \texttt{summary}=\texttt{False})$
    \EndIf
    \State \textbf{// Step 2: Send summarized context to next phase planner}
    \State \texttt{InitPlanner}($p_{i+1}$, \texttt{context}=$context$, \texttt{state}=$S$)
\EndFor
\end{algorithmic}
\end{small}
\end{algorithm}

\subsection{Generative Behavior and Execution}

\sloppy
xOffense supports three operational modes, specifically automatic, semi-automatic, and manual, with the automatic mode enabling fully autonomous testing.\allowbreak{} The \textit{Command Synthesizer} transforms TCG directives into precise, tool-specific instructions tailored to the target system and phase,\allowbreak{} addressing the challenge of accurate command generation.

For instance, a reconnaissance directive might yield \lstinline|nmap -sS -p 22,80,443 <target-ip>| for stealth scanning, while a scanning task might produce \lstinline|sqlmap -u http://<target-ip>/login --batch| to test for SQL injection vulnerabilities or \lstinline|nikto -h http://<target-ip>| for web server analysis. In the exploitation phase, commands like \lstinline|use exploit/windows/smb/ms17_010_eternalblue| in \textit{Metasploit}.

The \textit{Action Executor} runs these commands via a Python \textit{Paramiko}-based interactive shell on a Kali Linux environment, simulating human interactions with high fidelity. This component seamlessly processes tool-specific instructions generated by the \textit{Command Synthesizer}, enabling robust interaction with the target system through simulated keyboard operations. The \textit{Action Executor} is optimized for the \textbf{Qwen3-32B} model, which, despite its constrained 16,384-token context window, integrates the innovative \textit{MemAgent} \cite{yu2025memagent} framework to handle extended contexts effectively. Drawing on MemAgent’s segment-based processing and reinforcement learning (RL)-optimized memory mechanism, as described in the referenced study, the Action Executor processes arbitrarily long outputs by iteratively reading command results in chunks and updating a fixed-length memory. This approach ensures linear computational complexity, allowing xOffense to manage verbose tool outputs without performance degradation, even beyond the Qwen3-32B’s native context limit. To address the challenge of excessive or redundant output, a sophisticated filtering mechanism employs the MemAgent-enhanced LLM to extract critical information when results exceed 8,000 characters, preserving only actionable insights for analysis. These insights are relayed to the \textit{Task Orchestrator} for further processing, ensuring contextual coherence and minimizing computational overhead. By incorporating MemAgent’s ability to selectively retain relevant data while discarding distractors, the \textit{Action Executor} enhances the system’s efficiency and scalability, enabling robust pentest in resource-constrained environments and contributing to xOffense's capability to handle complex, long-context workflows with precision.

\section{Implementation, Benchmark Dataset and Metrics} \label{sec_experiment}
\label{sec:experiments}

To thoroughly assess the effectiveness and practicality of \textit{xOffense}, we designed a comprehensive experimental setup that evaluates not only task completion rates but also the scalability, adaptability, and efficiency of the system under realistic pentest conditions. This section details the experimental settings, models, fine-tuning method, benchmark datasets, evaluation metrics and presents an in-depth analysis of the empirical results.

\subsection{Experimental Setup}

\subsubsection{Attacker Environment} 
\label{sec:attacker_env}
All pentest experiments are executed from a dedicated attacker machine configured as a VMware virtual workstation running Kali Linux 2025 \cite{KaliLinux}. Kali is chosen for its comprehensive pentest toolkit and compatibility with industry-standard workflows. This virtualized attacker host is provisioned with 8 vCPUs, 16 GB RAM, and 120 GB storage, ensuring stable execution of both offensive tools and the \textit{xOffense} multi-agent framework within a single environment.

\subsubsection{Victim Environment} 
Two types of target environments are deployed corresponding to the benchmark datasets:
\begin{itemize}
    \item \textbf{AUTOPENBENCH} \cite{autopenbench}: Tasks are instantiated as Docker containers, which are hosted on a separate virtual machine to avoid resource contention with the attacker host. This victim VM is configured with 4 vCPUs, 8 GB RAM, and 80 GB storage, and placed in the same NAT network as the attacker machine to ensure direct connectivity.
    \item \textbf{AI-Pentest-Benchmark} \cite{aipentestbenchmark}: Vulnerable machines are directly imported from official VulnHub distributions and executed as VMware virtual machines without modification to their default specifications, in order to preserve the original exploitation conditions. All machines are assigned to the same NAT network as the attacker host to guarantee consistent communication.
\end{itemize}

% \subsubsection{Tool configuration} 
% \textcolor{blue}{bo sung viec cau hinh cac tool de khang dinh cac nghien cuu khac và xoffense dung chung phien ban nao....}
% For the baseline comparisons (e.g., VulnBot, PentestGPT), we report the performance metrics as documented in their recent comprehensive evaluations \cite{Kong2025VulnBot}. While xOffense is deployed on a modern Kali Linux 2025 environment, the baseline frameworks were evaluated on slightly earlier distributions (e.g., Kali 2023).
% To address potential concerns regarding tool-version bias, it is important to note the nature of the evaluation datasets. The benchmarks utilized (AutoPenBench and AI-Pentest-Benchmark) consist of static, fully isolated environments containing legacy vulnerabilities (e.g., CVEs spanning from 2014 to 2024). For such frozen targets, the execution logic, payloads, and standard outputs of the core toolchain (e.g., Nmap, Nikto, sqlmap, and Metasploit) remain highly deterministic and functionally invariant across these minor version updates. Therefore, the minor discrepancies in tool versions between the baseline environments and ours are not expected to introduce any significant bias, ensuring a fair comparative assessment of the underlying AI methodologies.

\subsubsection{Evaluation Protocol}

Within the unified environment described in Section~\ref{sec:attacker_env}, xOffense utilizes a core toolchain with specific versions detailed in Table~\ref{tab:tools}. These industry-standard utilities maintain stable command-line interfaces (CLI) and operational syntaxes, ensuring that the agent's reasoning logic remains functionally compatible with prior benchmark studies despite minor version iterations.

\begin{table}[h]
\centering
\small
\caption{Operational toolchain and versions (Kali Linux 2025)}
\label{tab:tools}
\begin{tabular}{ll}
\toprule
\textbf{Tool} & \textbf{Version} \\
\midrule
Nmap                    & 7.95 \\
Nikto                   & 2.5.0 \\
WPScan                  & 3.8.28 \\
sqlmap                  & 1.9.2\#stable \\
Metasploit Framework    & 6.4.50 \\
Hydra                   & 9.5 \\
Enum4linux              & 0.9.1 \\
Gobuster                & 3.8.2 \\
Dirb                    & 2.22 \\
\bottomrule
\end{tabular}
\end{table}

The evaluation itself follows a standardized protocol designed to guarantee fair and reproducible comparisons across all evaluated methods. Each agent is executed under identical constraints, including a maximum interaction budget of 5 steps per task, a timeout of 60 minutes per run, and a fixed token budget per model invocation. All methods are restricted to the same set of tools and environment configurations, without access to external resources beyond those explicitly defined.

Each experimental scenario is repeated for 5 independent runs with different random seeds to account for stochasticity in LLM outputs. Reported results are averaged across runs, and we additionally report standard deviations where applicable.

The evaluation follows a goal-oriented setting, where an episode is considered successful if the agent achieves the predefined exploitation objective within the allowed budget. Intermediate actions, such as reconnaissance and vulnerability identification, are not independently scored but contribute to overall task completion.

To prevent potential bias, all models are evaluated on the same subset of benchmark environments, and no task-specific tuning is performed during evaluation. Furthermore, we ensure that training data used for fine-tuning does not overlap with evaluation benchmarks to mitigate data leakage. This protocol is consistently applied to both the proposed framework and all baselines to ensure comparability and reliability of the reported results.

\subsection{LLM models}
The fine-tuned model, namely Qwen3-32B-finetune is hosted on a dedicated compute node equipped with an NVIDIA A100 GPU having 80 GB VRAM. The same hardware is also used during the fine-tuning process to accelerate training efficiency. For inference, the model is exposed via an API endpoint tunneled through \texttt{ngrok}, allowing the attacker machine to interact with the model as an external service. This separation ensured that LLM inference did not compete with pentest tasks for system resources, while also replicating realistic deployment conditions where models are often served remotely.

\subsubsection{Evaluated Baseline Models}
The experimental evaluation utilizes a broad spectrum of large language models to establish a rigorous comparative baseline. This selection encompasses leading proprietary models such as GPT-4o, alongside high-performance open-source architectures including Llama3.3-70B, Llama3.1-405B, and DeepSeek-V3. For the mid-scale category, Qwen3-32B-base serves as the foundational open-source representative, which is subsequently optimized into Qwen3-32B-finetune for specialized penetration testing tasks. 

Empirical consistency is maintained by enforcing a standardized decoding strategy across all tested models. The inference process is configured with a temperature of 0.5 and a top-p value of 0.9. This specific combination strikes a deliberate balance between the deterministic accuracy required for security tool command synthesis and the generative flexibility necessary for exploring complex attack paths. Furthermore, the top-k parameter is set to 40, allowing for a diverse yet controlled vocabulary selection that prevents the model from generating highly improbable tokens.
Each model interaction is allocated a maximum budget of 4096 tokens. This extended context window is essential for processing verbose security logs and generating comprehensive multi-stage exploit scripts without the risk of mid-sentence truncation. This universal hyperparameter setup isolates the architectural design and multi-agent orchestration of each framework as the primary variables driving the observed performance results.
% \textcolor{blue}{ngoai temperature can bo sung top-p, max tokens, decoding strategy....}

\subsubsection{Fine-Tuning Methodology}

Unlike conventional large-scale model adaptations, which demand significant computational overhead, \textit{xOffense} leverages LoRA to achieve domain specialization. LoRA reduces the parameter footprint by freezing the base model's weights and training only a compact set of adapter matrices. This approach dramatically lowers the number of trainable parameters by over 99\%, enabling efficient fine-tuning even on standard GPU infrastructures.

To further address the memory bottlenecks of handling a 32B-parameter model, we employ DeepSpeed ZeRO-3 \cite{rajbhandari2020zero} optimization. ZeRO-3 partitions model states, which includes optimizer, gradients, and parameters across multiple GPUs, achieving linear scalability. Additionally, FlashAttention v2 \cite{dao2022flashattention} is integrated to optimize attention computation, reducing memory usage and accelerating training by up to 3x compared to standard attention implementations. These combined techniques allow us to efficiently fine-tune Qwen3-32B for pentest workloads with a significant reduction in hardware demands.
And the fine-tuning dataset comprised two main corpora as follows:
\begin{itemize}
\item \textbf{PentestData}: This dataset is meticulously curated to encompass domain-specific question-answer pairs, each enriched with synthetically generated CoT reasoning traces. The reasoning steps are encapsulated within \verb|<think>| tags, enabling the model to learn structured, step-by-step logical deduction processes tailored for pentest scenarios. To construct PentestData, we aggregate and standardize write-ups from over \textbf{1,000 machines} across leading cybersecurity platforms, including TryHackMe~\cite{TryHackMe}, HackTheBox~\cite{HackTheBox}, and VulnHub~\cite{VulnHub}. These write-ups are systematically processed to extract task-specific reasoning paths, exploit procedures, and decision-making sequences relevant to offensive security operations. In addition, we incorporate supplementary pentest datasets from HuggingFace Datasets Hub~\cite{HuggingFaceDatasets}, focusing on cybersecurity knowledge bases, pentest techniques, and practical guides for commonly used security tools. This comprehensive integration ensures that PentestData serves as a robust and diverse resource for training models in autonomous pentest workflows.

\sloppy
\item \textbf{WhiteRabbitNeo}: A high-quality JSONL-formatted dataset comprising instruction-response pairs, specifically curated for cybersecurity tasks.\allowbreak{} Although the original dataset lacked explicit CoT reasoning annotations, it is standardized during preprocessing by appending empty \verb|<think>| tags to each sample.\allowbreak{} This structural unification ensures compatibility with CoT-augmented training pipelines and facilitates subsequent fine-tuning for step-by-step reasoning abilities.\allowbreak{} The dataset draws from real-world offensive and defensive cybersecurity scenarios, encompassing exploitation techniques, payload crafting, and red-team/blue-team interactions, sourced from the WhiteRabbitNeo community contributions \cite{WhiteRabbitNeo}.

\end{itemize}

\subsubsection{Prompt Setting}

To ensure methodological transparency and reproducibility, \textit{xOffense} adopts a \textbf{single system-wide initialization prompt} (\textit{init prompt}) that governs the full penetration workflow, rather than using distinct prompt policies for individual agents. This unified prompt defines the global objective, operational boundaries, execution environment assumptions, output contracts, and cross-phase memory usage, thereby enforcing consistent reasoning behavior throughout reconnaissance, scanning, and exploitation.

Concretely, the init prompt specifies four mandatory control dimensions: (i) \textbf{mission context} (target objective and scope constraints), (ii) \textbf{state continuity} (reuse of prior successful actions and shell status), (iii) \textbf{actionability requirements} (tool-compatible and directly executable commands), and (iv) \textbf{structured output formatting}. The unified design reduces prompt-policy drift across phases and improves execution stability when the planner performs iterative update and reflection.

To reduce parsing ambiguity, the init prompt enforces strict output contracts: planning outputs must be serialized within \verb|<json></json>| tags, while executable commands must be enclosed in \verb|<execute></execute>| tags. This explicit interface is essential because downstream modules parse model outputs programmatically. In addition, long execution logs are summarized before being re-injected into context, which mitigates context dilution and helps maintain coherent decision-making over multi-step attack trajectories.

The prompt follows a grey-box principle: it does not expose implementation internals, but injects selective operational state, including prior successful or failed tasks, shell continuity indicators, and compressed history summaries. This strategy preserves cross-step coherence without overwhelming the model with raw terminal traces.

For all experiments, prompt execution used a unified inference policy to isolate the effect of architecture and prompting from decoding variance. Specifically, we use temperature $=0.5$, top-$p=0.9$, top-$k=40$, and a maximum generation budget of 4096 tokens per interaction, consistent across compared models. In the deployed \textit{xOffense} implementation, hidden chain-of-thought traces (when produced inside \verb|<think>| tags) are removed before downstream processing, ensuring evaluation is based only on actionable outputs while retaining stable external behavior.

For reproducibility, a representative structured output schema governed by the init prompt is shown below:

\begin{lstlisting}[language={},caption={Representative structured planning output used in xOffense prompts.},label={lst:prompt_plan_schema}]
<json>
[
    {
        "id": "1",
        "dependent_task_ids": [],
        "instruction": "Enumerate open services on 10.10.10.5 with version detection.",
        "action": "Shell"
    },
    {
        "id": "2",
        "dependent_task_ids": ["1"],
        "instruction": "Assess HTTP service on 10.10.10.5:80 for common web vulnerabilities.",
        "action": "Web"
    }
]
</json>
\end{lstlisting}

\subsection{Benchmark Datasets}

To ensure a rigorous and practically relevant evaluation, we selected two complementary benchmarks that cover both synthetic vulnerabilities and realistic multi-stage exploitation scenarios. Together, these benchmarks provide a balanced testbed for assessing the adaptability and robustness of automated pentest systems.

\paragraph{AutoPenBench} 
This benchmark defines a total of 33 pentest tasks, spanning both instructional “in-vitro” exercises and real-world CVEs. 
The in-vitro set (22 tasks) reflects fundamental vulnerability classes frequently highlighted in industry rankings such as the OWASP Top 10, including weak access control (such as misconfigured sudo, world-writable shadow files), web application flaws (such as path traversal, SQL injection, file upload RCE), and insecure network configurations (such as SNMP misconfiguration, ARP spoofing). 
In addition, four cryptography tasks evaluate resilience against improper or weak cryptographic implementations, such as brute-forcing Diffie–Hellman keys. 
Beyond these educational tasks, the benchmark incorporates 11 real-world CVEs ranging from 2014 to 2024 with CVSS scores between 7.5 and 10.0. 
These include critical vulnerabilities widely recognized for their impact and prevalence, such as Log4Shell (CVE-2021-44228), Heartbleed (CVE-2014-0160), SambaCry (CVE-2017-7494), and Spring4Shell (CVE-2022-22965). 
By combining foundational categories with critical CVEs, AutoPenBench provides a structured yet realistic environment to evaluate whether agents can transition from basic exploitation to handling high-severity vulnerabilities that have historically dominated real-world attack campaigns.

\paragraph{AI-Pentest-Benchmark.} 
While AutoPenBench focuses on containerized vulnerabilities, the AI-Pentest-Benchmark evaluates AI agents on complete end-to-end exploitation workflows across 13 real-world vulnerable machines drawn from VulnHub. 
These machines are categorized by difficulty into easy (such as Victim1, Library2, Funbox, WestWild), medium (such as Cengbox2, Devguru, Symfonos2), and hard (such as Insanity, TempusFugit). 
Each machine defines a structured set of reconnaissance, exploitation, privilege escalation, and general technique subtasks, amounting to 152 tasks in total. 
The vulnerabilities embedded within these machines reflect common pentest scenarios, including web application flaws (SQL injection, XSS, CSRF/SSRF), network service weaknesses (FTP/AD enumeration, brute-force authentication), code-level issues (deserialization, command injection), and post-exploitation techniques (cronjob analysis, misconfigured system files, privilege escalation via user access exploitation). 
Notably, many of these tasks map directly to recurring weakness categories in the CWE Top 25 and OWASP Top 10, ensuring that success on this benchmark corresponds to capabilities relevant in practical offensive security operations. 
The benchmark is particularly challenging because the ultimate goal is to achieve root access on each machine, requiring coherent reasoning across reconnaissance, exploitation, and privilege escalation stages. 
Previous studies have shown that even large-scale proprietary models such as GPT-4o and Llama3.1-405B are unable to achieve root-level compromise without human assistance, underscoring the difficulty and realism of this benchmark.

% \subsection{Evaluation Metrics}

% Performance was primarily measured through:
% \begin{itemize}
% \item \textbf{Overall Task Completion Rate}: The percentage of target machines successfully compromised within predefined step limits.
% \item \textbf{Sub-task Completion Rate}: A granular metric that evaluates the success rate of individual subtasks (e.g., service enumeration, vulnerability scanning), providing insights into the system's robustness at each procedural stage.
% \end{itemize}

\subsection{Evaluation Metrics}

To evaluate the performance of \textit{xOffense}, we employ three complementary metrics that capture both high-level task success and fine-grained sub-task robustness.

\subsubsection{Overall Task Completion Rate}
This metric measures the percentage of target machines successfully compromised like the obtained flag within the allowed interaction budget. Formally:
\[
\text{Overall Rate} = \frac{\#\text{ compromised machines}}{\#\text{ total machines}}
\]
This provides a coarse-grained view of whether an agent can achieve end-to-end exploitation across categories such as Access Control (AC), Web Security (WS), Network Security (NS), Cryptography (CRPT), and Real-world tasks.

\subsubsection{Sub-task Completion Rate (1 Experiment)}
To gain insight into intermediate stages of pentest, we evaluate sub-task success rates, including service enumeration and vulnerability detection. Each benchmark defines a set of subtasks $S$. A sub-task is considered successful if it is completed in at least one of the five independent runs:
\[
\text{Subtask-1Exp} = \frac{|\{s \in S \mid \exists i \in [1,5],\, \text{success}(s,i)\}|}{|S|}
\]
This metric highlights the agent's ability to eventually solve a sub-task, even if not consistently across all runs.

\subsubsection{Sub-task Completion Rate (5 Experiments)}
To measure robustness and consistency, we also compute the cumulative completion rate across all five runs. In this case, we count the total number of successful sub-tasks over all experiments and normalize by the maximum possible number of successes:
\[
\text{Subtask-5Exp} = \frac{\sum_{i=1}^{5} \text{ successes}(i)}{5 \times |S|}
\]
This stricter metric rewards agents that not only succeed once but can repeatedly complete subtasks across independent executions.

Together, these three metrics provide a balanced view: (i) overall penetration capability, (ii) eventual solvability of subtasks, and (iii) robustness of performance under repeated trials.

\section{Evaluation and results} \label{sect_evaluation_n_results}
\subsection{Experiment Scenarios}

We design five experimental scenarios to comprehensively evaluate \textit{xOffense} across synthetic and real-world settings:

\begin{itemize}
    \item \textbf{Scenario 1:} Overall task completion on AutoPenBench, measuring full machine compromise across AC, WS, NS, and CRPT categories. 
    \item \textbf{Scenario 2:} Sub-task completion on AutoPenBench (1 Experiment), where a sub-task is successful if solved in at least one of five runs. 
    \item \textbf{Scenario 3:} Sub-task completion on AutoPenBench (5 Experiments), aggregating successful subtasks across all five runs to capture consistency. 
    \item \textbf{Scenario 4:} Real-world exploitation on AI-Pentest-Benchmark without RAG, using six representative VulnHub machines. 
    \item \textbf{Scenario 5:} Real-world exploitation on AI-Pentest-Benchmark with RAG, highlighting the contribution of retrieval to complex exploitation chains. 
\end{itemize}

\subsection{Task Completion Performance Across pentest Categories}

\subsubsection{Overall Task Completion Performance}

% GHI CHÚ: Sử dụng môi trường 'table*' để bảng trải dài trên cả hai cột.
% LaTeX sẽ tự động đặt nó ở đầu trang tiếp theo. Đây là hành vi tiêu chuẩn.
% Tiêu đề cột được định dạng bằng \makecell để tự động ngắt dòng.
\begin{table*}[t!]
\centering
\caption{Overall Task Completion Rate on Target Machines. Our fine-tuned model demonstrates superior performance, especially in AC, NS, and Real-world categories.}
\label{tab:overall_completion}
\small 
% Thêm '|' để tạo đường kẻ dọc và sử dụng '\hline' cho đường kẻ ngang
\begin{tabular}{|l|c|c|c|c|c|c|}
\hline
\textbf{Category} & \textbf{GPT-4o} & \makecell{\textbf{Llama3.3-70B} \\ \textbf{(VulnBot)}} & \makecell{\textbf{Llama3.1-405B} \\ \textbf{(VulnBot)}} & \makecell{\textbf{Llama3.1-405B} \\ \textbf{(PentestGPT)}} & \makecell{\textbf{Qwen3-32B} \\ \textbf{(Base)}} & \makecell{\textbf{Qwen3-32B-finetune} \\ \textbf{(Ours)}} \\
\hline
AC & 1 (20.00\%) & 1 (20.00\%) & 3 (60.00\%) & 1 (20.00\%) & 2 (40.00\%) & \textbf{5 (100.00\%)} \\
\hline
WS & 2 (28.57\%) & 1 (14.29\%) & 2 (28.57\%) & 0 (0.00\%) & 2 (28.57\%) & \textbf{5 (71.42\%)} \\
\hline
NS & 3 (50.00\%) & 2 (33.33\%) & 2 (33.33\%) & 2 (33.33\%) & 3 (50.00\%) & \textbf{5 (83.33\%)} \\
\hline
CRPT & 0 (0.00\%) & 0 (0.00\%) & 0 (0.00\%) & 0 (0.00\%) & 0 (0.00\%) & \textbf{3 (75.00\%)} \\
\hline
Real-world & 1 (9.09\%) & 2 (18.18\%) & 3 (27.27\%) & 0 (0.00\%) & 3 (27.27\%) & \textbf{6 (54.54\%)} \\
\hline
\textbf{ALL} & 7 (21.21\%) & 6 (18.18\%) & 10 (30.30\%) & 3 (9.09\%) & 10 (30.30\%) & \textbf{24 (72.72\%)} \\
\hline
\end{tabular}
\end{table*}

% Bảng cho subtask 1 experiment (dữ liệu từ slide 18) với kẻ ô
\begin{table*}[t!]
\centering
\caption{Sub-task Completion Rate (1 Experiment) . Qwen3-32B-finetune shows the highest completion rate across all categories.}
\label{tab:subtask_1_exp}
\small
\begin{adjustbox}{width=1.0\linewidth}
\begin{tabular}{|l|c|c|c|c|c|c|c|}
\hline
\textbf{Category} & \makecell{\textbf{Llama3.3-70B}\\\textbf{(VulnBot)}} & \makecell{\textbf{Llama3.1-405B}\\\textbf{(VulnBot)}} & \makecell{\textbf{Llama3.3-70B}\\\textbf{(Base)}} & \makecell{\textbf{Llama3.1-405B}\\\textbf{(Base)}} & \makecell{\textbf{Llama3.1-405B}\\\textbf{(PentestGPT)}} & \makecell{\textbf{Qwen3-32B}\\\textbf{(Base)}} & \makecell{\textbf{Qwen3-32B-finetune}\\\textbf{(Ours)}} \\
\hline
AC   & 25 (11.90\%)  & 31 (14.76\%) & 16 (7.62\%)  & 21 (10.00\%) & 20 (9.52\%)  & 26 (8.20\%)  & \textbf{46 (14.51\%)}  \\
\hline
WS   & 24 (11.43\%)  & 30 (14.29\%) & 22 (10.48\%) & 26 (12.38\%) & 18 (8.57\%)  & 28 (8.83\%)  & \textbf{38 (11.98\%)}  \\
\hline
NS   & 12 (5.71\%)   & 11 (5.24\%)  & 10 (4.76\%)  & 9 (4.29\%)   & 6 (2.86\%)   & 11 (3.47\%)  & \textbf{15 (4.73\%)}   \\
\hline
CRPT & 15 (7.14\%)   & 18 (8.57\%)  & 17 (8.10\%)  & 18 (8.57\%)  & 12 (5.71\%)  & 22 (6.94\%)      & \textbf{38 (11.98\%)}  \\
\hline
Real-world & 49 (23.33\%)  & 55 (26.19\%) & 29 (13.81\%) & 29 (13.81\%) & 28 (13.33\%) & 79 (24.92\%) & \textbf{114 (35.96\%)} \\
\hline
\textbf{ALL} & 125 (59.52\%) & 145 (69.05\%)& 94 (44.76\%) & 103 (49.05\%)& 84 (40.00\%) & 166 (52.36\%)      & \textbf{251 (79.17\%)} \\
\hline
\end{tabular}
\end{adjustbox}
\end{table*}

% Bảng cho subtask 5 experiments (dữ liệu từ slide 19) với kẻ ô
\begin{table*}[t!]
\centering
\caption{Sub-task Completion Rate (5 Experiments) . Our model maintains a significant lead, demonstrating robustness and consistency.}
\label{tab:subtask_5_exp}
\small
\begin{adjustbox}{width=1.0\linewidth}
\begin{tabular}{|l|c|c|c|c|c|c|c|}
\hline
\textbf{Category} & \makecell{\textbf{Llama3.3-70B}\\\textbf{(VulnBot)}} & \makecell{\textbf{Llama3.1-405B}\\\textbf{(VulnBot)}} & \makecell{\textbf{Llama3.3-70B}\\\textbf{(Base)}} & \makecell{\textbf{Llama3.1-405B}\\\textbf{(Base)}} & \makecell{\textbf{Llama3.1-405B}\\\textbf{(PentestGPT)}} & \makecell{\textbf{Qwen3-32B}\\\textbf{(Base)}} & \makecell{\textbf{Qwen3-32B-finetune}\\\textbf{(Ours)}} \\
\hline
AC   & 87 (8.29\%)   & 107 (10.19\%) & 46 (4.38\%)  & 61 (5.81\%)  & 27 (2.57\%)  & 60 (3.78\%)  & \textbf{212 (14.51\%)} \\
\hline
WS   & 106 (10.10\%) & 116 (11.05\%) & 83 (7.90\%)  & 66 (6.29\%)  & 40 (3.81\%)  & 70 (4.42\%)  & \textbf{173 (10.91\%)} \\
\hline
NS   & 41 (3.90\%)   & 40 (3.81\%)   & 36 (3.43\%)  & 22 (2.10\%)  & 15 (1.43\%)  & 10 (0.63\%)  & \textbf{71 (4.67\%)}  \\
\hline
CRPT & 65 (6.19\%)   & 75 (7.14\%)   & 68 (6.48\%)  & 44 (4.19\%)  & 43 (4.10\%)  & 70 (4.42\%)      & \textbf{176 (11.10\%)} \\
\hline
Real-world & 166 (15.81\%) & 186 (17.71\%) & 99 (9.43\%)  & 67 (6.38\%)  & 56 (5.33\%)  & 155 (9.78\%) & \textbf{334 (21.07\%)} \\
\hline
\textbf{ALL} & 465 (44.29\%) & 524 (49.90\%) & 332 (31.62\%)& 260 (24.76\%)& 181 (17.24\%)& 365 (23.03\%)      & \textbf{966 (60.94\%)} \\
\hline
\end{tabular}
\end{adjustbox}
\end{table*}

Table~\ref{tab:overall_completion} presents the overall task completion rates across all evaluated models on the AutoPenBench dataset. The fine-tuned \textbf{Qwen3-32B-finetune} model achieved a remarkable \textbf{72.72\%} completion rate, substantially outperforming both its base variant, \textbf{Qwen3-32B-base} (30.30\%), and other state-of-the-art models, including GPT-4o (21.21\%), Llama3.1-405B (Paper) (30.30\%), and PentestGPT (9.09\%).

The performance disparity between \textbf{Qwen3-32B-finetune} and its base version is particularly noteworthy. Despite having identical model architecture and parameter size (32 billion parameters), the domain-specific fine-tuning enabled a \textbf{2.4x improvement} in task completion. This validates the effectiveness of our lightweight LoRA fine-tuning pipeline in adapting general-purpose models to specialized pentest workflows.

In the \textbf{AC} category, \textbf{Qwen3-32B-finetune} achieved a \textbf{100\% success rate}, a stark contrast to the 40.00\% of Qwen3-32B-base and the 60.00\% of Llama3.1-405B (Paper). Similarly, in \textbf{NS}, our model achieved \textbf{83.33\%} completion, surpassing all baselines, including Llama3.3-70B (33.33\%) and Qwen3-32B-base (50.00\%). Notably, even in the complex \textbf{Real-world} category, \textbf{Qwen3-32B-finetune} attained a \textbf{54.54\%} success rate, outperforming Qwen3-32B-base (27.27\%) and PentestGPT (0.00\%).

These findings demonstrate that fine-tuning on domain-relevant CoT data and incorporating robust task orchestration mechanisms can enable a quantized, resource-efficient model to match and even surpass the capabilities of larger, general-purpose LLMs in specialized scenarios. The consistent outperformance across categories further validates the robustness of our fine-tuning strategy, especially given the compute-efficient AWQ quantization.

\subsubsection{Sub-task Completion Performance (1 Experiment)}

To assess finer-grained capabilities, we evaluated sub-task completion in a single-run experiment (Table~\ref{tab:subtask_1_exp}). The term "1 Experiment" refers to the overall
sub-task completion rate across five experiments, where a sub-task is considered successful if it succeeds in at least one
experiment. \textbf{Qwen3-32B-finetune} achieved a \textbf{79.17\%} sub-task completion rate, outperforming Llama3.1-405B (Paper) (69.05\%) by a margin of \textbf{10.12\%}. This margin is significant, particularly when considering that Llama3.1-405B is a much larger model (405B parameters) operating in its native configuration.

In the \textbf{Real-world} category, \textbf{Qwen3-32B-finetune} achieved a \textbf{35.96\%} sub-task completion rate, more than doubling that of Qwen3-32B-base (24.92\%) and outperforming Llama3.1-405B (Paper) (26.19\%). Similarly, in the CRPT category, the fine-tuned model demonstrated a \textbf{3.41\%} improvement over Llama3.1-405B (Paper).

Interestingly, while Qwen3-32B-base achieved moderate results (52.36\%), its performance gap to Qwen3-32B-finetune (79.17\%) illustrates the critical role of domain adaptation. The base model, though capable of handling general security tasks, struggled with multi-step reasoning and contextual coherence, particularly in chained exploit scenarios. The fine-tuned model’s superior performance confirms that its CoT-driven prompt alignment and RAG-assisted knowledge retrieval mechanisms provide a tangible advantage in executing complex task sequences.

\subsubsection{Sub-task Completion Performance (5 Experiments)}

To evaluate robustness and stability, we conducted aggregated experiments over five runs (Table~\ref{tab:subtask_5_exp}). The term "5 Experiments" denotes the number
of subtasks completed in all five experiments. \textbf{Qwen3-32B-finetune} maintained its lead with a \textbf{60.94\%} sub-task completion rate, significantly outperforming Llama3.1-405B (Paper) (49.90\%) and Qwen3-32B-base (23.03\%). This robustness is critical in practical pentesting workflows, where variance due to environmental noise and complex task dependencies often degrades model performance.

In the \textbf{AC} category, \textbf{Qwen3-32B-finetune} achieved a remarkable \textbf{14.51\%}, which is \textbf{4.32\% higher} than Llama3.1-405B (Paper). For \textbf{WS}, our model reached \textbf{10.91\%}, outperforming all baselines by a significant margin. Notably, even in categories where task chains are inherently volatile, such as \textbf{Real-world}, the fine-tuned model achieved \textbf{21.07\%}, compared to 17.71\% for Llama3.1-405B (Paper) and only 9.78\% for Qwen3-32B-base.

While a performance drop of approximately \textbf{18\%} from the single-experiment run to the aggregated runs was observed, this is expected due to increased task complexity and stochastic failures inherent in autonomous pentesting. Nevertheless, the fine-tuned model’s consistency across these iterations underscores its robustness, particularly when contrasted with PentestGPT’s 17.24\% sub-task completion in the same setting.

\subsubsection{Comparative Insights}

A critical observation from these experiments is the disproportionate performance leap achieved through fine-tuning relative to model size. Despite being a 32B parameter model, \textbf{Qwen3-32B-finetune} consistently outperformed larger counterparts like Llama3.1-405B (405B parameters) across every evaluation metric. This validates our hypothesis that task orchestration, RAG-driven context augmentation, and parameter-efficient tuning techniques (LoRA + ZeRO-3 + FlashAttention) can bridge, and in specialized scenarios, exceed the performance gap traditionally associated with sheer model size.

Furthermore, the disparity between Qwen3-32B-base and Qwen3-32B-finetune exemplifies the inadequacy of using general-purpose LLMs in specialized pentesting workflows without domain adaptation. The base model, though architecturally identical, lacked the reasoning depth and context-coherence required for intricate attack path planning, resulting in lower task and sub-task completion rates.

\subsection{Evaluation on Complex Real-World Exploitation Chains}
\subsubsection{Performance without RAG (No-RAG)}
To assess the baseline capabilities of our proposed system \textit{xOffense} in realistic offensive security scenarios, we conducted experiments on the same set of six real-world vulnerable machines as utilized in the VulnBot \cite{Kong2025VulnBot} evaluation: \textbf{Victim1}, \textbf{Library2}, \textbf{Sar}, \textbf{WestWild}, \textbf{Symfonos2}, and \textbf{Funbox}. This machine set, originally derived from the AI-Pentest-Benchmark, covers a diverse range of exploitation challenges, including misconfigurations, weak authentication, remote code execution, privilege escalation, and multi-step attack chains. By adopting this identical set, we ensure methodological consistency and enable a direct, fair comparison with prior work.

\begin{figure*}[t]
\centering
\includegraphics[width=\textwidth]{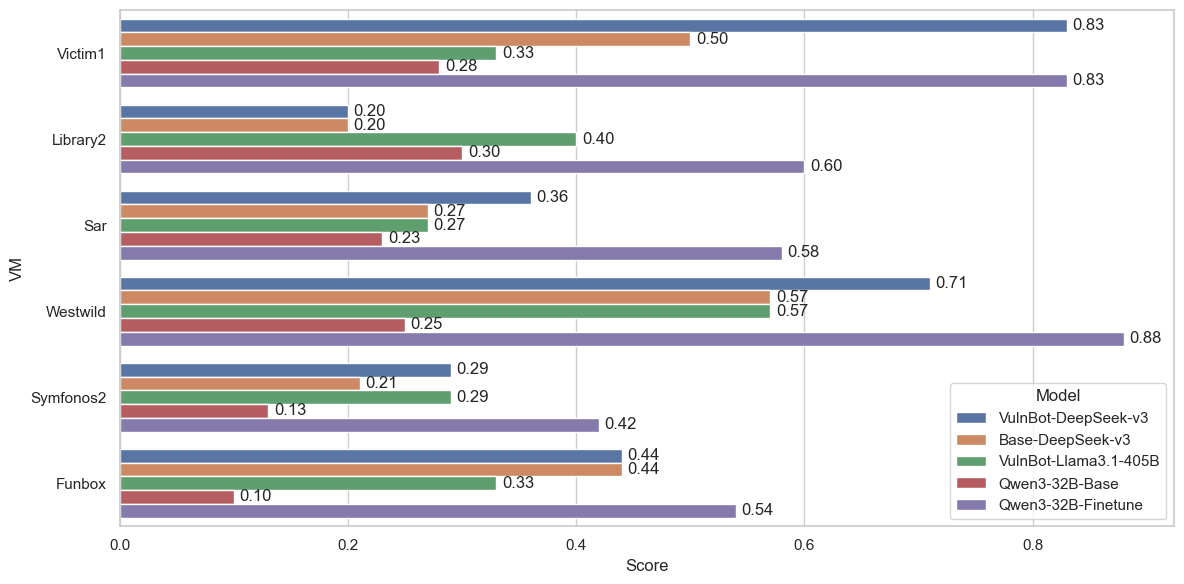}
\caption{Comparison of sub-task completion rates across six real-world vulnerable machines in a No-RAG setting.}
\label{fig:ai-pentest-no-rag}
\end{figure*}

The experiments were conducted in a fully autonomous mode, without any human intervention or RAG support. Each target machine was tested in five independent runs, and the reported performance represents the best sub-task completion rate per machine, following the AI-Pentest-Benchmark scoring methodology. Figure~\ref{fig:ai-pentest-no-rag} presents the comparative results across multiple models, including \textbf{VulnBot-Llama3.1-405B}, \textbf{VulnBot-DeepSeek-v3}, their respective base models, and our proposed \textbf{Qwen3-32B} variants (base and finetuned).

The results reveal several noteworthy patterns. First, \textbf{Qwen3-32B-finetune} consistently surpasses its base counterpart across all six machines, with particularly significant improvements on \textit{Victim1} (+0.55), \textit{Library2} (+0.30), and \textit{WestWild} (+0.63). These gains highlight the effectiveness of domain-specific fine-tuning in strengthening the model’s exploitation reasoning and procedural robustness. Second, while VulnBot-DeepSeek-v3 remains highly competitive, achieving the highest score on \textit{Victim1} (0.83) and \textit{WestWild} (0.71), our fine-tuned Qwen3-32B achieves comparable or superior performance on most other machines, including leading results on \textit{Sar} (0.58) and \textit{Funbox} (0.54).

Notably, performance disparities are strongly correlated with the complexity of exploitation chains. Targets such as \textit{Symfonos2} and \textit{Funbox}, which demand multi-stage privilege escalation and exploitation of non-trivial service configurations, clearly benefit from the enhanced contextual reasoning introduced via fine-tuning. This observation underscores the critical role of model specialization in addressing the inherent unpredictability and diversity of real-world pentest environments. In summary, the No-RAG evaluation confirms that \textbf{xOffense-Qwen3-32B-finetune} can autonomously achieve competitive, and in some cases state-of-the-art, performance in realistic offensive security scenarios, even without external retrieval augmentation. This establishes a robust performance baseline for subsequent RAG-enhanced evaluations.

% \begin{figure}[H]
%     \centering
%     \includegraphics[width=1\linewidth]{images/ai-pentest-no-rag-hori.png}
%     \caption{Comparison of subtask completion rates across six real-world vulnerable machines in a No-RAG setting.}
%     \label{fig:ai-pentest-no-rag}
% \end{figure}

\begin{figure*}[!t]
    \centering
    \includegraphics[width=\textwidth]{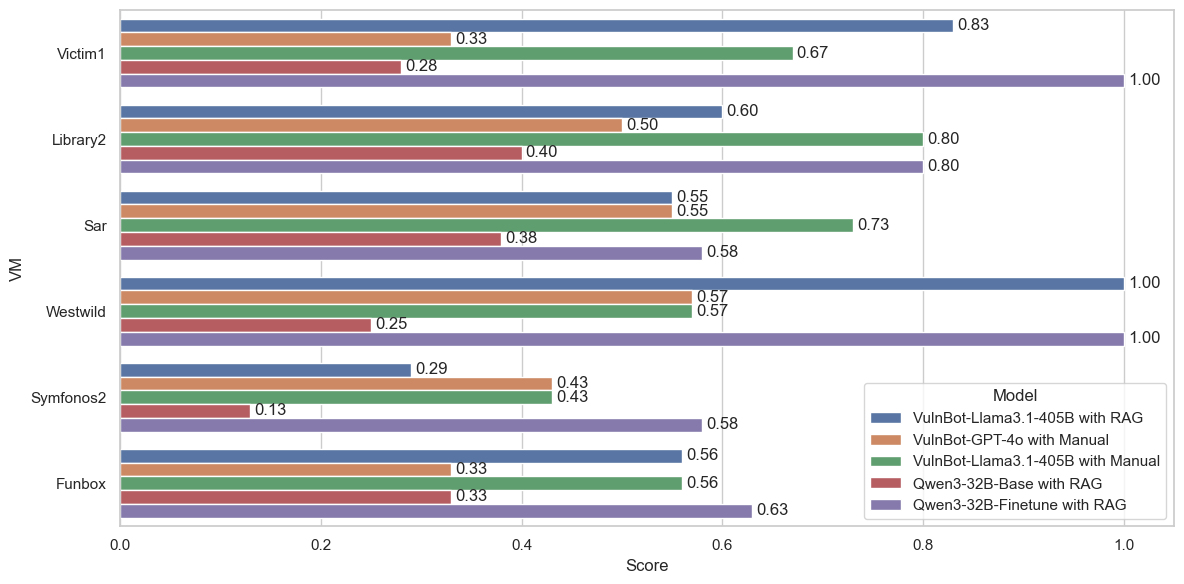}
    \caption{Comparison of sub-task completion rates across six real-world vulnerable machines with RAG setting.}
    \label{fig:ai-pentest-with-rag}
\end{figure*}

\subsubsection{Performance with RAG (RAG)}

When augmenting the evaluation with the Knowledge Repository module, a substantial shift in performance trends emerges across the six real-world exploitation targets \emph{(see Fig.~\ref{fig:ai-pentest-with-rag})}. Compared to the baseline (No-RAG), the Qwen3-32B-Finetune model demonstrates marked improvement, achieving perfect completion scores on Victim1 and WestWild (1.00) and notable gains on Library2 (+0.20) and Symfonos2 (+0.16). Similarly, moderate increases are observed for Sar and Funbox, reflecting the model’s enhanced capability to navigate multi-step attack chains when supported by targeted, contextually relevant prior knowledge.

The gains are less pronounced for Qwen3-32B-Base, with performance remaining comparatively low on challenging targets such as Symfonos2 (0.13) and WestWild (0.25). This disparity underscores the role of fine-tuning in maximizing the benefits of retrieval augmentation, as without alignment to domain-specific exploitation strategies, the retrieved information alone is insufficient to ensure consistent execution success. When compared against VulnBot baselines, Qwen3-32B-Finetune with RAG achieves competitive or superior results in four out of six targets, matching the best baseline performance on Library2 (0.80) and surpassing it on Victim1 and WestWild. This suggests that RAG integration not only mitigates the limitations of the base model but also allows the fine-tuned variant to close the gap, or in certain scenarios, outperform human-assisted frameworks.

These improvements can be attributed to three key factors: (1) the retriever’s ability to surface high-relevance exploitation procedures from a curated cybersecurity corpus, (2) the fine-tuned model’s capacity to integrate external information into coherent multi-step reasoning, and (3) the reduction of hallucination-driven dead ends, which are particularly detrimental in constrained exploitation environments. Collectively, these findings reinforce the notion that RAG is a critical enabler for scalable, high-fidelity automated pentest in complex real-world settings.

\section{Threats to Validity} \label{sec_threats}

\subsection{Internal validity}
The fine-tuning of Qwen3-32B on a CoT-enriched pentest dataset introduces potential internal threats. Certain vulnerability classes and exploitation strategies are disproportionately represented, which may bias the model toward specific attack vectors while limiting its capacity to generalize to underrepresented scenarios. Moreover, the integration of prompting strategies and toolchains may embed implicit task-specific heuristics, raising the possibility that reported improvements partly reflect dataset artifacts rather than genuine reasoning ability. Such factors must be considered when interpreting performance gains on structured benchmarks.  

\subsection{External validity}
The evaluation settings of AutoPenBench and AI-Pentest-Benchmark, which approximate realistic penetration workflows, yet cannot fully capture the heterogeneity of production-scale environments. Operational networks often exhibit greater variability in topology, non-standard configurations, active defenses, and deception mechanisms that remain absent from current benchmarks. In addition, adversarial tactics evolve over time, whereas benchmarks are necessarily static. Consequently, the generalizability of results to enterprise systems, heterogeneous infrastructures, or zero-day exploitation scenarios should be regarded with caution.  

\subsection{Construct validity}
Task completion rate and exploitation success were employed as primary evaluation metrics. While suitable for quantifying functional effectiveness, these measures neglect other dimensions that are central to pentest practice. Attributes such as stealth, efficiency of resource utilization, time-to-compromise, and resilience against detection are critical to operational realism yet remain unaccounted for in the adopted benchmarks. Furthermore, binary success measures fail to capture partial progress or incremental compromise, potentially obscuring nuances in agent behavior across complex exploitation chains.  

\subsection{Reliability}
The reproducibility of results may be affected by stochastic factors inherent in both large language model inference and auxiliary system tools. Hardware variation, runtime conditions, network latency, and nondeterministic outputs from scanning utilities can yield divergent agent behaviors even under identical inputs. Standardized configurations and repeated trials mitigate these effects but do not eliminate them entirely, implying that replications across platforms or over extended periods may observe non-negligible variance.

In sum, although the reported findings provide strong evidence of the capabilities of xOffense, these validity concerns underscore the need for broader empirical validation. Expanding evaluations to encompass more diverse infrastructures, adversarially adaptive defenses, and richer performance metrics would strengthen the robustness, scalability, and practical applicability of autonomous pentest systems. 

%%%% bo sung
\section{Ethical Considerations and Responsible Use} \label{sec_ethicalconsideration}

This work addresses autonomous penetration testing, a domain with inherent dual-use implications. While the proposed xOffense framework is intended to support defensive cybersecurity practices, it may also be subject to misuse if deployed without appropriate safeguards. The system is designed strictly for authorized security assessment, research, and training purposes in controlled environments, such as benchmark platforms and laboratory settings, and must only be used with explicit legal authorization from system owners. It is not intended for unauthorized exploitation of real-world systems.

Due to its capability to automate multi-stage penetration testing workflows, the framework may lower the barrier for conducting offensive operations, including large-scale vulnerability discovery and exploitation. This highlights the importance of responsible usage and user awareness when interacting with such systems. Users are expected to adhere to applicable legal frameworks, institutional policies, and professional ethical standards when deploying or extending the proposed approach.

To mitigate potential risks, all experiments in this study are conducted strictly within sandboxed and publicly available benchmark environments, including AutoPenBench and AI-Pentest-Benchmark, without interaction with live production systems or undisclosed vulnerabilities. In practical deployments, additional safeguards are necessary, including restricting execution to authorized and monitored environments, enforcing logging and auditing of system actions, limiting tool usage and external network access, and incorporating human-in-the-loop validation for high-impact decisions. Future implementations should further integrate policy-based constraints and access control mechanisms to prevent unintended or malicious usage.

Beyond its offensive capabilities, this line of research also motivates the development of more robust defensive mechanisms. Understanding how autonomous agents identify and exploit vulnerabilities can inform the design of improved intrusion detection systems, adaptive defenses, and secure system architectures. As such, this work contributes not only to offensive security automation but also to advancing defensive cybersecurity research.

This research adheres to responsible cybersecurity and AI research practices. The study does not involve zero-day exploitation or unauthorized system access, and all evaluations are conducted on simulated or explicitly permitted targets. Future extensions should follow coordinated disclosure principles and comply with relevant legal and ethical standards, in line with established guidelines for responsible AI and cybersecurity research.

\section{Conclusion and Future Work} \label{sec_conclusion}

This work presented xOffense, an independent, fully autonomous multi-agent framework for pentest, designed to address persistent limitations in existing systems such as context loss, limited reasoning continuity, and dependence on large proprietary models. By integrating a fine-tuned, mid-scale open-source LLM (Qwen3-32B) with a novel grey-box phase prompting mechanism and a purpose-built orchestration architecture, xOffense achieves accurate multi-stage decision-making and robust tool integration across the entire pentest lifecycle.

Our evaluation on \textit{AutoPenBench} and \textit{AI-Pentest-Benchmark} demonstrated that xOffense consistently outperforms both larger commercial LLMs, such as GPT-4o and LLaMA3-70B, and leading open-source baselines, such as PentestGPT and VulnBot-LLaMA3-405B. The framework attained an overall task completion rate of 72.72\% and a sub-task completion rate of up to 79.17\%, while successfully exploiting complex real-world targets. These results highlight that a domain-specialized, mid-scale model, when paired with targeted reasoning guidance, can match or exceed the capabilities of state-of-the-art large-scale systems, offering a cost-effective and reproducible solution for autonomous offensive security operations.

Future work will explore three main directions. First, we aim to optimize the command generation module, potentially through structured function calling, to further improve execution precision. Second, we plan to enhance the robustness of long-running process handling and strengthen the retrieval-augmented generation mechanism with automated updates from vulnerability intelligence sources such as ExploitDB and GitIngest. Third, we intend to extend xOffense’s capabilities to support advanced web and GUI interactions via browser automation, enabling it to tackle a broader range of pentest scenarios.

\bibliographystyle{unsrt}
\bibliography{refs.bib}

\end{document}